\documentclass[sn-mathphys-num]{sn-jnl}% Basic Springer Nature Reference Style/Chemistry Reference Style

\usepackage{graphicx}%
\usepackage{multirow}%
\usepackage{amsmath,amssymb,amsfonts}%
\usepackage{amsthm}%
\usepackage{mathrsfs}%
\usepackage[title]{appendix}%
\usepackage{xcolor}%
\usepackage{textcomp}%
\usepackage{manyfoot}%
\usepackage{mathrsfs}
\usepackage{booktabs}%
\usepackage{algorithm}%
\usepackage{algorithmicx}%
\usepackage{algpseudocode}%
\usepackage{listings}%

\theoremstyle{thmstyleone}%
\theoremstyle{thmstyletwo}%

\theoremstyle{thmstylethree}%
\begin{document}

\title[Article Title]{An Adolescent and Near-Resonant Planetary System Near the End of Photoevaporation}

\author*[1,2,3]{\fnm{Mu-Tian} \sur{Wang}}\email{mutianwang@smail.nju.edu.cn}
\author*[3]{\fnm{Fei} \sur{Dai}}\email{fdai@hawaii.edu}
\author*[1,2]{\fnm{Hui-Gen} \sur{Liu}}\email{huigen@nju.edu.cn}
\author[4,5]{\fnm{Howard} \sur{Chen}}
\author[6]{\fnm{Zhecheng} \sur{Hu}}
\author[7]{\fnm{Erik} \sur{Petigura}}
\author[8]{\fnm{Steven} \sur{Giacalone}}
\author[9]{\fnm{Eve} \sur{Lee}}
\author[10]{\fnm{Max} \sur{Goldberg}}
\author[11,12]{\fnm{Adrien} \sur{Leleu}}
\author[13]{\fnm{Andrew W.} \sur{Mann}}
\author[13]{\fnm{Madyson G.} \sur{Barber}}
\author[14]{\fnm{Joshua N.} \sur{Winn}}
\author[15]{\fnm{Karen A.} \sur{Collins}}
\author[15]{\fnm{Cristilyn N.} \sur{Watkins}}
\author[15]{\fnm{Richard P.} \sur{Schwarz}}
\author[15]{\fnm{Howard M.} \sur{Relles}}
\author[16]{\fnm{Francis P.} \sur{Wilkin}}
\author[17,18]{\fnm{Enric} \sur{Palle}}
\author[17,18]{\fnm{Felipe} \sur{Murgas}}
\author[19]{\fnm{Avi} \sur{Shporer}}
\author[20]{\fnm{Ramotholo} \sur{Sefako}}
\author[21]{\fnm{Keith} \sur{Horne}}
\author[22,23]{\fnm{Hugh P.} \sur{Osborn}}
\author[12,22]{\fnm{Yann} \sur{Alibert}}
\author[24]{\fnm{Luca} \sur{Fossati}}
\author[12,22]{\fnm{Andrea} \sur{Fortier}}
\author[25]{\fnm{Sérgio} \sur{Sousa}}
\author[26]{\fnm{Alexis} \sur{Brandeker}}
\author[27]{\fnm{Pierre} \sur{Maxted}}
\author[16]{\fnm{Alexia} \sur{Goldenberg}}

\affil[1]{\orgdiv{School of Astronomy and Space Science}, 
          \orgname{Nanjing University}, 
          \orgaddress{\city{Nanjing}, \postcode{210023}, \country{China}}}

\affil[2]{\orgdiv{Key Laboratory of Modern Astronomy and Astrophysics, Ministry of Education}, 
          \orgname{Nanjing University}, 
          \orgaddress{\city{Nanjing}, \postcode{210023}, \country{People’s Republic of China}}}

\affil[3]{\orgdiv{Institute for Astronomy}, 
          \orgname{University of Hawai‘i}, 
          \orgaddress{\street{2680 Woodlawn Drive}, \city{Honolulu}, \state{HI}, \postcode{96822}, \country{USA}}}

\affil[4]{\orgdiv{Department of Aerospace, Physics, and Space Sciences}, 
          \orgname{Florida Institute of Technology}, 
          \orgaddress{\city{Melbourne}, \state{FL}, \postcode{32901}, \country{USA}}}

\affil[5]{\orgdiv{Sellers Exoplanet Environments Collaboration (SEEC)}, 
          \orgname{NASA Goddard Space Flight Center}, 
          \orgaddress{\city{Greenbelt}, \state{MD}, \postcode{20771}, \country{USA}}}

\affil[6]{\orgdiv{Department of Astronomy}, 
          \orgname{Tsinghua University}, 
          \orgaddress{\city{Beijing}, \postcode{10084}, \country{China}}}

\affil[7]{\orgdiv{Department of Physics \& Astronomy}, 
          \orgname{University of California Los Angeles}, 
          \orgaddress{\city{Los Angeles}, \state{CA}, \postcode{90095}, \country{USA}}}

\affil[8]{\orgdiv{Department of Astronomy}, 
          \orgname{California Institute of Technology}, 
          \orgaddress{\city{Pasadena}, \state{CA}, \postcode{91125}, \country{USA}}}

\affil[9]{\orgdiv{Department of Astronomy \& Astrophysics}, 
          \orgname{University of California, San Diego}, 
          \orgaddress{\city{La Jolla}, \state{CA}, \postcode{92093-0424}, \country{USA}}}

\affil[10]{\orgdiv{Laboratoire Lagrange}, 
           \orgname{Université Côte d’Azur, CNRS, Observatoire de la Côte d’Azur}, 
           \orgaddress{\street{Boulevard de l’Observatoire}, \city{Nice Cedex 4}, \postcode{06304}, \country{France}}}

\affil[11]{\orgdiv{Observatoire astronomique}, 
           \orgname{Université de Genève}, 
           \orgaddress{\street{Chemin Pegasi 51}, \city{Versoix}, \postcode{1290}, \country{Switzerland}}}

\affil[12]{\orgdiv{Space Research and Planetary Sciences, Physics Institute}, 
           \orgname{University of Bern}, 
           \orgaddress{\street{Gesellschaftsstrasse 6}, \city{Bern}, \postcode{3012}, \country{Switzerland}}}

\affil[13]{\orgdiv{Department of Physics and Astronomy}, 
           \orgname{The University of North Carolina at Chapel Hill}, 
           \orgaddress{\city{Chapel Hill}, \state{NC}, \postcode{27599}, \country{USA}}}

\affil[14]{\orgdiv{Department of Astrophysical Sciences}, 
           \orgname{Princeton University}, 
           \orgaddress{\street{4 Ivy Lane}, \city{Princeton}, \state{NJ}, \postcode{08544}, \country{USA}}}

\affil[15]{\orgdiv{Center for Astrophysics | Harvard \& Smithsonian}, 
           \orgaddress{\street{60 Garden Street}, \city{Cambridge}, \state{MA}, \postcode{02138}, \country{USA}}}

\affil[16]{\orgdiv{Department of Physics and Astronomy}, 
           \orgname{Union College}, 
           \orgaddress{\street{807 Union St.}, \city{Schenectady}, \state{NY}, \postcode{12308}, \country{USA}}}

\affil[17]{\orgname{Instituto de Astrofísica de Canarias (IAC)}, 
           \orgaddress{\city{La Laguna, Tenerife}, \postcode{38205}, \country{Spain}}}

\affil[18]{\orgdiv{Departamento de Astrofísica}, 
           \orgname{Universidad de La Laguna (ULL)}, 
           \orgaddress{\city{La Laguna, Tenerife}, \postcode{38206}, \country{Spain}}}

\affil[19]{\orgdiv{Department of Physics and Kavli Institute for Astrophysics and Space Research}, 
           \orgname{Massachusetts Institute of Technology}, 
           \orgaddress{\city{Cambridge}, \state{MA}, \postcode{02139}, \country{USA}}}

\affil[20]{\orgname{South African Astronomical Observatory}, 
           \orgaddress{\street{P.O. Box 9, Observatory}, \city{Cape Town}, \postcode{7935}, \country{South Africa}}}

\affil[21]{\orgdiv{SUPA Physics and Astronomy}, 
           \orgname{University of St. Andrews}, 
           \orgaddress{\city{Fife}, \postcode{KY16 9SS}, \country{United Kingdom}}}

\affil[22]{\orgdiv{Center for Space and Habitability}, 
           \orgname{University of Bern}, 
           \orgaddress{\street{Gesellschaftsstrasse 6}, \city{Bern}, \postcode{3012}, \country{Switzerland}}}

\affil[23]{\orgdiv{Department of Physics}, 
           \orgname{ETH Zurich}, 
           \orgaddress{\street{Wolfgang-Pauli-Strasse 2}, \city{Zurich}, \postcode{CH-8093}, \country{Switzerland}}}

\affil[24]{\orgname{Space Research Institute, Austrian Academy of Sciences}, 
           \orgaddress{\street{Schmiedlstrasse 6}, \city{Graz}, \postcode{A-8042}, \country{Austria}}}

\affil[25]{\orgname{Instituto de Astrofisica e Ciencias do Espaco, Universidade do Porto}, 
           \orgaddress{\street{CAUP, Rua das Estrelas}, \city{Porto}, \postcode{4150-762}, \country{Portugal}}}

\affil[26]{\orgdiv{Department of Astronomy}, 
           \orgname{Stockholm University}, 
           \orgaddress{\street{AlbaNova University Center}, \city{Stockholm}, \postcode{10691}, \country{Sweden}}}

\affil[27]{\orgdiv{Astrophysics Group}, 
           \orgname{Keele University}, 
           \orgaddress{\street{Lennard Jones Building}, \city{Staffordshire}, \postcode{ST5 5BG}, \country{United Kingdom}}}

\abstract{
Young exoplanets provide vital insights into the early dynamical and atmospheric evolution of planetary systems. Many multi-planet systems younger than 100 Myr exhibit mean-motion resonances, likely established through convergent disk migration. Over time, however, these resonant chains are often disrupted, mirroring the Nice model proposed for the Solar System. We present a detailed characterization of the $\sim$200-Myr-old TOI-2076 system, which contains four sub-Neptune planets between 1.4 and 3.5 Earth radii. We demonstrate that its planets are near but not locked in mean-motion resonances, making the system dynamically fragile. The four planets have comparable core masses but display a monotonic increase in hydrogen and helium (H/He) envelope mass fractions (stripped-1\%-5\%-5\%) with decreasing stellar insolation. This trend is consistent with atmospheric mass-loss due to photoevaporation, which predicts that the envelopes of irradiated planets either erode completely or stabilize at a residual level of $\sim$1\% by mass within the first few hundred million years, with more distant, less-irradiated planets retaining most of primordial envelopes. Additionally, previous detections of metastable helium outflows rule out a pure water-world scenario for TOI-2076 planets. Our finding provides direct observational evidence that the dynamical and atmospheric reshaping of compact planetary systems begin early, offering an empirical anchor for models of their long-term evolution.
}

%\keywords{keyword1, Keyword2, Keyword3, Keyword4}

\maketitle
\section{Main}
A key prediction of modern planet formation theory is that planets migrate as they form in a protoplanetary disk, usually towards the central star \citep{TerquemPapaloizou2007}. In multi-planet systems, differential migration continuously alters orbital period ratios, often leading to capture into mean-motion resonances (MMR,\cite{MurrayDermott}). When period ratios approach integer commensurate (e.g. 3:2, 2:1), mutual interaction between planets add up coherently, thereby locking pairs or even chains of planets into resonances. At the disk’s inner edge, a reversal in migration torque halts this process, parking resonant chains at short orbital periods of a few days, giving rise to the compact ($<1$ AU), multi-planet systems commonly discovered by the {\it Kepler} mission \cite{Fabrycky2014}.

Recent discoveries of young multi-planet systems provide strong empirical support for this framework: resonant configurations are prevalent among young ($<$100-Myr-old), close-in planetary systems, accounting for $\sim80\%$ of the observed sample \cite{Dai2024,Hamer2024}. Moreover, the fraction of resonant planets steadily declines with age dropping to about 30\% for adolescent systems (0.1-1Gyr) and further down to 15\% for mature systems ($>$1Gyr, \cite{huang2023and}). The prevailing explanation is that, after disk dispersal, the loss of eccentricity and inclination damping renders resonant configurations dynamically unstable and lead to close encounters \cite{Izidoro2017,Goldberg_stability}. Such a dynamical dissolution of resonant configuration mirrors the scenario proposed in the Nice model for the early Solar System \cite{Nesvorny2018}.

\subsection*{Photodynamical Model of TOI-2076}

We present a thorough characterization of TOI-2076 system \cite{Hedges2021,Barber2025}, whose adolescent age of $\sim210\pm20$ Myr (constrained by ensemble study of comoving stars \cite{Barber2025}) makes it a key signpost for studying dynamical evolution and the erosion of primordial atmospheres. TOI-2076 is a K-dwarf hosting four sub-Neptune planets (e, b, c, d) with orbital periods spanning 3 to 35 days. The outer three planets (b, c, d) reside near mean-motion resonances with period ratios of $P_c/P_b =2.03$ (near $2:1$) and $P_d/P_c =1.671$ (near $5:3$). Their proximity to resonance can be quantified by the parameter $\Delta = \frac{P_2/P_1}{j/(j-k)} - 1$, for a pair near a $j$:$j-k$ resonance. For TOI-2076, we measure $\Delta_{bc} = 1.5\%$ and $\Delta_{cd} = 0.3\%$, values that are broadly consistent with the offsets typically seen in \emph{Kepler} near-resonant systems ($\Delta \simeq 1$–$2\%$; \cite{Fabrycky2014}). The innermost planet e is discovered more recently \cite{Barber2025}, which is far from MMR and likely dynamically detached from the other planets. The whole system is likely well-aligned with the host star's rotation axis \cite{Frazier}.

Transit timing variations (TTVs, Figure~\ref{fig:ttv}a-d) provide a powerful probe of gravitational interactions in multi-planet systems near MMR, and in turn constrain their masses and other orbital parameters \cite{Lithwick2012_ttv}.
Planets b and c exhibit the most prominent TTVs, characterized by anti-correlated sinusoids with amplitudes of $\sim$10 minutes and a dominant periodicity of 720 days. 720 days closely matches the theoretical "super-period": $P_s = 1/|j/P_{\rm 2}-(j-k)/P_{\rm 1}|$ \cite{Lithwick2012_ttv}, and is indicative of near-resonant dynamics. In contrast, in-resonance, librating TTVs occur with the libration period $P_l \approx P_{\rm orb} (\frac{m_p}{m_\star})^{-2/3}$ \cite{Nesvorny2016}. In the near-resonant regime, TTVs constrain only the product of planetary mass and eccentricity \cite{Lithwick2012_ttv}, which is evident in our TTV retrievals (Methods and ED Figure 1).

We therefore employed a photodynamical model to break the mass–eccentricity degeneracy. A photodynamical model performs direct N-body integration of a planetary system while jointly modeling photometric and radial velocity (RV) data. (`Photodynamical Modeling' in Methods). Moreover, modeling RVs of young planet hosts poses additional challenges due to surface magnetic activity, which induces quasi-periodic variability at the stellar rotation period (approximately 50 m/s) which may overwhelm planetary signals ($\lesssim2.5$ m/s) (ED Figure 2). We incorporated a Gaussian Process (GP) regression in our photodynamical model ('Radial Velocity Analysis' in Methods). The rotation-induced stellar activity can also induce photometry activity, therefore the GP hyperparameters for RV analysis were first trained on photometric variability in the {\it TESS} light curve and then applied to the combined dataset. 

With measured masses of $M_e=4.7\pm1.5$ M$_\oplus$, $M_b=6.8\pm1.8$ M$_\oplus$, $M_c=7.2\pm1.4$ M$_\oplus$, and $M_d=7.3\pm2.7$ M$_\oplus$, we found that all four TOI-2076 planets have comparable masses below the threshold for runaway gas accretion \cite{Rafikov}. 
These masses place them in the super-Earth to sub-Neptune regime, as shown in the mass–radius diagram ($R<4R_\oplus$, Figure~\ref{fig:ttv}i) along with theoretical composition curves \cite{Chen2016,Zeng2019,Aguichine2024}. The innermost planet, TOI-2076 e (3-day orbit), appears rocky, with its H/He envelope completely stripped off. The outer three near-resonant planets are lower in density and should contain volatile-rich atmosphere \cite{Leleu2024}. TOI-2076 b (10-day, $R = 2.59 \pm 0.1$ R$_\oplus$) is consistent with having a rocky core and a thin ($\sim$1\% by mass) H/He envelope. TOI-2076 c and d (20- and 35-day orbits, $R_c = 3.54 \pm 0.14$ R$_\oplus$, $R_d = 3.27 \pm 0.13$ R$_\oplus$) likely retain more massive envelopes ($\sim$5\%), thanks to their larger core masses and lower insolation levels.

\subsection*{Near Resonant and Metastable}

To demonstrate the dynamical state of TOI-2076 planets near MMR, we mapped the posterior orbital solutions of TOI-2076 onto the Hamiltonian framework of resonant dynamics \cite{Nesvorny2016}. For first-order $k:(k-1)$ MMR (2:1 for TOI-2076 b and c), the Hamiltonian takes the form:

\begin{align}
    H = &-\frac{GM_*m_1}{2a_1}-\frac{GM_*m_2}{2a_2} - \frac{Gm_1 m_2}{a_2}\\    \nonumber
        &\times[fe_1\cos(k\lambda_2-(k-1)\lambda_1-\varpi_1) \\                 \nonumber
        &+ge_2\cos(k\lambda_2-(k-1)\lambda_1-\varpi_1)]
\end{align}
where $m,e,\lambda,\varpi,$ and $a$ are the planet mass, eccentricity, mean longitude, longitude of pericenter, and semi-major axis. $f$ and $g$ are order-of-unity constants that vary with the specific resonance $k$ \cite{MurrayDermott}. The first two terms describe the Keplerian motion of the planets, while the third captures their mutual gravitational interactions. Through a series of canonical transformations, this Hamiltonian can be reduced to a simplified form \cite{Nesvorny2016},
\begin{equation}
\label{eq:first_order_mmr}
    H = -(\Psi - \delta)^2 - \sqrt{2\Psi} \cos \psi
\end{equation}
the resonant angle, $\psi=k\lambda_2-(k-1)\lambda_1 -\hat{\varpi}$, and its conjugate action $\Psi$ form the canonical pair describing resonant dynamics. The action $\Psi$ scales with a linear combination of the planets’ eccentricities, while the parameter $\delta$ quantifies the system’s proximity to exact resonance and governs the phase-space topology of the resonant Hamiltonian, which we show in Figure~\ref{fig:near_resonant}a. Formally resonant, librating solutions enclosed by separatrices emerge only when $\delta > 0.95$. Here, the resonant angle librates (oscillates) with small amplitude around the stable equilibrium point offset from the origin. A representative example is the TOI-1136 planets (orange points) which lie deep within the separatrix and exhibit libration \cite{Dai2023}. In contrast, TOI-2076 b and c have $\delta = -7.0 \pm 0.8$, placing them well outside the 2:1 resonance. In this regime, the Hamiltonian has not bifurcated, no separatrix exists, and the resonant angle circulates freely through the full $2\pi$ range (Figure~\ref{fig:near_resonant}b).

For second-order $k:(k-2)$ MMR (e.g. 5:3 for TOI-2076 c and d), the approximate form of the resonant interaction is \cite{Hadden2019}:
\begin{equation}
    H = \frac{1}{8}A(J-J^*)^2-\Tilde{\epsilon}J \cos(\theta)
\end{equation}
where $A$ and $\Tilde{\epsilon}$  depend on the resonance index $k$ and planet-to-star mass ratios \cite{Hadden2019}. The action-angle pair $J$ and $\theta$ plays a similar role to $(\Psi, \psi)$ in the first-order case, and $J^*$ measures the proximity to resonance.  The Hamiltonian of second-order MMR is plotted in Figure~\ref{fig:near_resonant}c. Truly resonant, librating orbits bounded by separatrices emerge only when $J^*>0$.  For the TOI-2076 c–d pair, all posterior samples yield $J^* < 0$, placing the system firmly outside the 5:3 resonance and exhibiting circulating resonant arguments (Figure~\ref{fig:near_resonant}d). In contrast, the TOI-1136 e-f pair is deep in 7:5 second-order MMR and is likely in a librating state \cite{Dai2023}.

If TOI-2076 had preserved the pristine orbital configuration established by convergent disk migration, its planets would be expected to reside in mean-motion resonances, characterized by librating resonant angles and period ratios close to exact commensurabilities ($\Delta < 0.1\%$; \cite{Keller2025}). In contrast, our dynamical analysis suggests that the system has modestly diverged by $\Delta \sim 1\%$ from what would have been expected from the simplest picture of disk migration, and it currently occupies a near-resonant, circulating state. Such architectures may result from a range of dynamical processes, operating either at the time of formation or over the course of their subsequent evolution, including stochastic perturbations from disk turbulence during migration \cite{Batygin2015_capture}, the divergent torques produced by a receding inner disk edge \cite{Liu2017}, interactions with residual planetesimals \cite{Wu2024}, or orbital changes induced by early atmospheric erosion \cite{Wang_lin_2023}. TOI-2076 does not conform to the simplest expectations of disk migration, yet this is also true for many near-resonant \emph{Kepler} systems. Owing to its well-constrained dynamical state and young age, TOI-2076 offers a valuable benchmark for testing different mechanisms that produce such enigmatic architectures.

The libration of the resonant angle is critical for the long-term orbital stability. A canonical example is the 3:2 mean-motion resonance between Neptune and Pluto, where the resonant angle $\phi_{\rm NP} = 2\lambda_{N}-3\lambda_{P}+\varpi_P$ librates around 180$^\circ$. This libration ensures that whenever Pluto approaches perihelion, Neptune is safely 90$^\circ$ ahead or behind in its orbit, thereby avoiding close encounters despite their crossing orbits \cite{Williams}. If the resonant angle were circulating, the system would explore a broader range of configurations including those permitting close encounters, and thereby increasing the risk of dynamical instability \cite{Obertas2017, Rath2022, Lammers2024}. Fortunately, the orbital eccentricities of the TOI-2076 planets are low $e\simeq0.01$ resulting in minimal overlap of nearby MMRs \cite{Wisdom1980}, and the system is not in immediate danger of orbital instability \cite{Hu2025}. However, TOI-2076 remains vulnerable to further dynamical excitation, such as planetesimal scattering \cite{Wu2024} or secular interaction with undetected outer companions \cite{Lithwick2011}. The near-resonant configuration may be disrupted by close encounters in next 10s-100s of Myr \cite{Dai2024}. This is reminiscent of the proposed early instabilities in the Solar System in the Nice Model \cite{Nesvorny2018}. Early dynamical evolution may be common and play a fundamental role in sculpting out the final architectures of planetary systems.

\subsection*{Erosion of Primordial Atmosphere}

Sub-Neptune planets are thought to form with substantial primordial H/He envelopes \cite{Lee2015}, yet the radii of mature, close-in small planets exhibit bimodal distributions \cite{Fulton}, with a lower peak ($1-1.8R_\oplus$) representing super-Earths and rocky planets without primordial (H/He) envelopes, and a larger radius peak ($1.8-3.5R_\oplus$) corresponding to mini-Neptunes that retain thin ($\sim$1\%-by-mass) H/He or water-rich envelopes \cite{Owen2017,Luque_water}. One avenue to produce this feature is through atmospheric mass-loss process driven by a combination of residual formation heat \cite{Ginzburg} and X-ray and ultraviolet (XUV) driven photoevaporation \cite{Owen2017}.

\begin{figure}[!ht]
	\centering
	\includegraphics[width=1.0\textwidth]{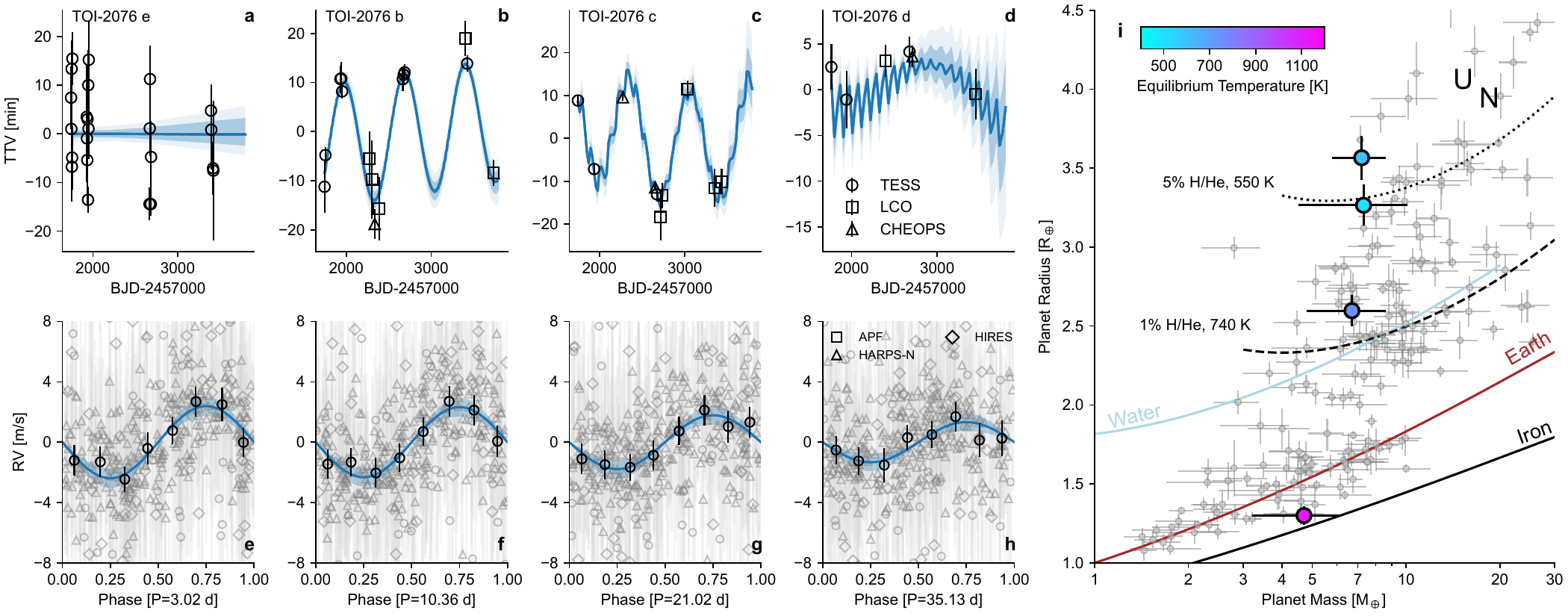}
	\caption{\textbf{Transit timing variations and radial velocity observation, and measured masses and radii of TOI-2076 planets.}
		(\textbf{A-D}) Transit timing variations of TOI-2076 planets, observations from TESS, LCO, and CHEOPS are denoted as open circles, squares, and triangles. The median and 1$\sigma$ interval of photodynamical posterior are shown as blue solid lines and shaded area. 
            (\textbf{E-H}) phase-folded radial velocity data (triangle: HARPS-N, diamond: HIRES, square: Automated Planet Finder) of TOI-2076 planets, with stellar activity signals removed. The solid line and shaded area are the median and 1$\sigma$ interval of RV posterior. Black errorbars are binned RV data in 8 evenly-spaced bins over a planetary period. 
            (\textbf{I}) The masses and radii of TOI-2076 planets from joint photodynamics and RV model are shown in circles, colored by their equilibrium temperatures. Grey circles are exoplanets from the Exoplanet Archive (as of 2025-05-23) with mass/radius constraints better than 4$\sigma$.  Mass-radius relations with different compositions are overplotted \cite{Chen2016,Zeng2019,Aguichine2024}.}
	\label{fig:ttv} 
\end{figure}

\begin{figure}[!ht]
	\centering
	\includegraphics[width=1.0\textwidth]{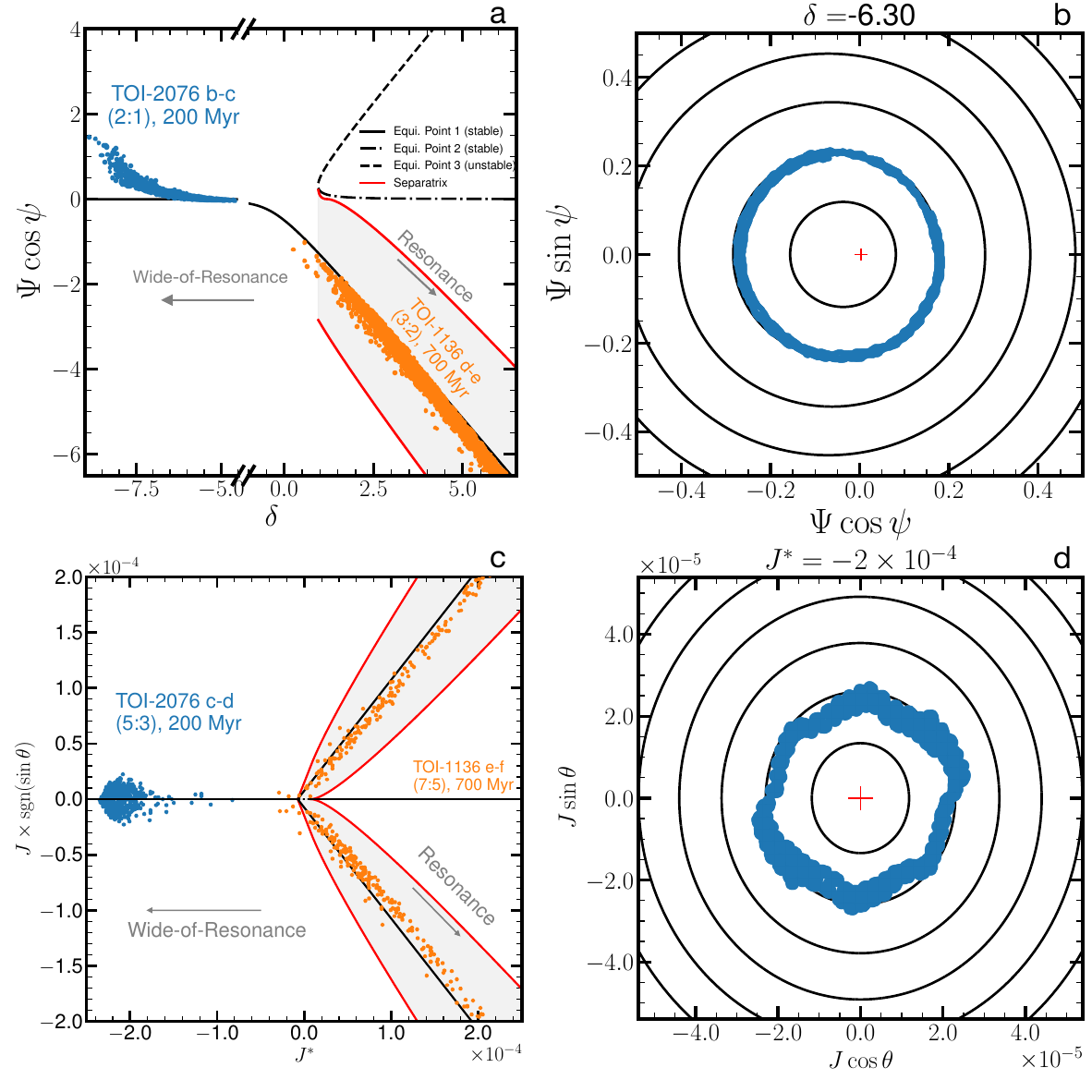}
	\caption{\textbf{Mean-motion resonance diagram of TOI-2076 planets.}
		(\textbf{A}) The Hamiltonian map of first-order mean-motion resonance \cite{Nesvorny2016}. Resonant regime is shown as the gray area. The 2:1 TOI-2076 b-c pair (blue points) lies beyond the resonant regime.
        (\textbf{B}) The integrated orbit of TOI-2076 b-c on the action-angle phase space at $\delta=-6.3$. The trajectory encloses the origin, suggesting TOI-2076 b-c has a circulating resonant angle. 
        (\textbf{C}) The second-order mean-motion resonance Hamiltonian map \cite{Hadden2019}. Resonance regime is shown in gray. The 5:3 TOI-2076 c-d (blue points) is near-resonant while the 7:5 TOI-1136 e-f (orange points) lies in the true second-order resonant regime.
        (\textbf{D}) The integrated orbits of TOI-2076 c-d on the action-angle phase space at $J^*=-2\times10^{-4}$, which also shows the TOI-2076 c-d have a circulating resonant angle.
        }
	\label{fig:near_resonant}
\end{figure}

TOI-2076 presents a rare opportunity to probe the timescale of the emergence of planet radius bimodal distributions. Metastable helium outflows detected from all three outer planets \cite{Zhang2023} provide evidence against a pure water world scenario and suggest at least some amount of H/He is present in the envelope and is susceptible to XUV-driven photoevaporation. The near-resonant orbits and low eccentricities of planets b, c, and d indicate they have not experienced giant impacts which may also erode atmosphere \cite{Biersteker}. Theoretical studies show that the gas accretion rate onto planetary cores is primarily regulated by the core mass and the opacity of the envelope, with a secondary dependence on the local disk conditions set by formation location \citep{Lee2015}. Given the similar planetary core masses constrained from TTV observation, the TOI-2076 planets likely formed with comparable primordial H/He atmospheres from the protoplanetary disk (`Initial Gas Envelope Contents' in Method). While at 200-Myr-old, the planetary envelope contents have diverged (Figure~\ref{fig:ttv}i), likely caused by the photoevaporation effect due to the elevated XUV flux from the young host star \cite{Ribas2005}. 

\begin{figure}[!ht]
	\centering
	\includegraphics[width=1.0\textwidth]{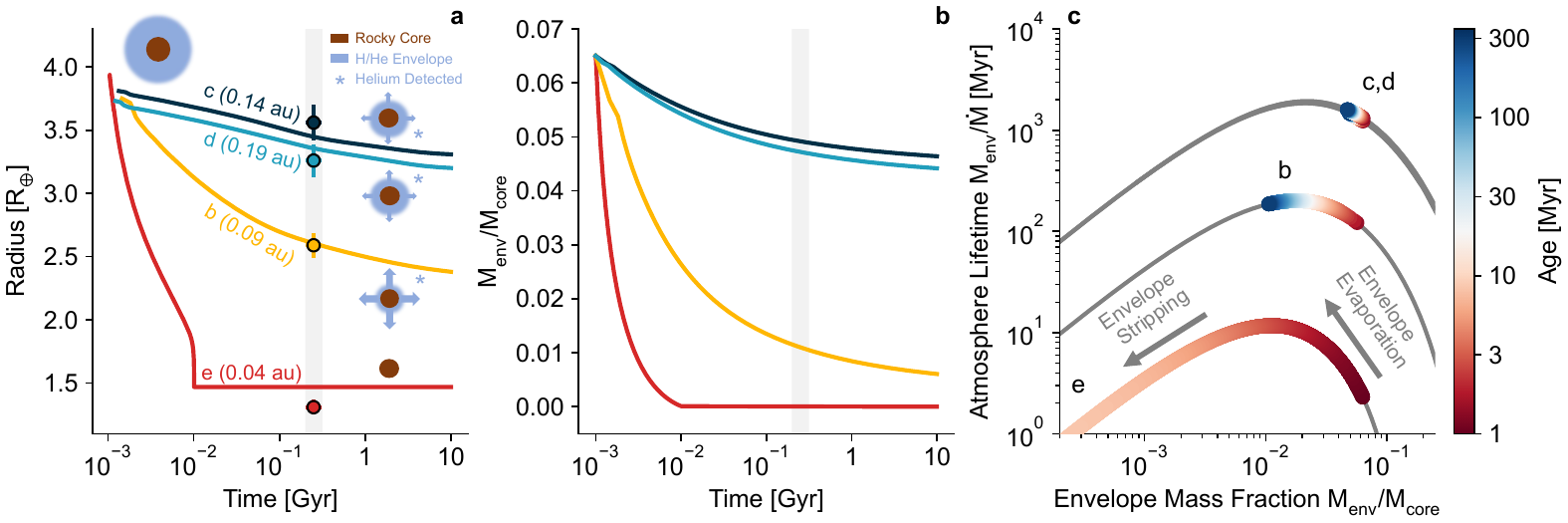} 
	\caption{
    \textbf{XUV-driven photoevaporation mass-loss evolution of TOI-2076 planets.}
    (\textbf{A}) The radius evolution of and TOI-2076 e (red), b (yellow), c (dark blue), and d (light blue). The errorbars show the median and 68\% credible interval of observed planetary radii, and the vertical gray band show the system age $\sim200$-Myr-old.
    (\textbf{B}) The evolution of H/He envelope mass fraction $M_{\rm env}/M_{\rm core}$.
    (\textbf{C}) The TOI-2076 planets atmosphere lifetime, defined as the envelope mass divided by mass loss rate $\dot M$. The $M_{\rm env}/M_{\rm core}$ evolution tracks from the simulation are interpolated on the evolutionary track and colored by the time.
    }
	\label{fig:atm_evol}
\end{figure}

Figure~\ref{fig:atm_evol}a presents our simulated atmospheric evolution for the TOI-2076 planets. Our model includes Kelvin-Helmholtz contraction and XUV-driven photoevaporation (`Evolution of H/He Evolution' in Methods). Planet e is included in the simulations with the caveat that its history may have been altered by a giant impact\cite{Biersteker}, or early mass-loss during disk dispersal \cite{Owen_boiloff}. Orbital migration due to tidal interaction and mass-loss is likely limited \cite{Lithwick_repulsion,Wang_lin_2023,Hanf}, we therefore assume that the planets remain static throughout the photoevaporation process. To highlight their differential evolution, the simulation begins with all four planets having the same H/He envelope mass fraction of 6.5\% (Figure~\ref{fig:atm_evol}b). Intense early XUV irradiation at the 3-day orbit causes TOI-2076 e to lose its entire atmosphere within the first 10 Myr. TOI-2076 b also underwent rapid atmospheric erosion, losing most of its envelope within the first 100 Myr but stabilizing around 1\% by mass. The outer planets TOI-2076 c and d experience only modest mass loss. Our simulation suggests the outer three planets are still undergoing mass loss, consistent with observed metastable helium outflows from all three outer planets \cite{Zhang2023}.

The observed differential atmospheric loss in TOI-2076 can be best understood through their diverging atmospheric lifetimes compared to system age, defined as the envelope mass divided by the current mass-loss rate (Figure~\ref{fig:atm_evol}c). For XUV-driven photoevaporation, this timescale scales approximately as $\propto M_p^{1.4} a^{2.1}$ \cite{Owen2017}, where $M_p$ is the planet mass and $a$ is the semi-major axis. TOI-2076 e, being the least massive and closest to the star, has the shortest atmospheric lifetime of tens of Myr, thus it is completely stripped at its current age. Planets with a heavier H/He envelope are more resistant to mass loss. However, there is a turning point when H/He envelope reaches about 1\% by mass. Here, the atmosphere is massive enough that it can no longer be treated with a thin-atmosphere assumption. Its scale height is comparable to the radius of the rocky core, increasing the cross section to XUV flux and shortening the atmospheric lifetime. This produces a peak in lifetime around $\sim$1\% envelope mass. In systems that have been sculpted by photoevaporation, planets are either fully stripped (super-Earth, left corner of Figure~\ref{fig:atm_evol}c) or they retain $\sim1\%$-by-mass H/He envelopes (mini-Neptunes, near the peak of Figure~\ref{fig:atm_evol}c). 

\begin{figure}[!ht]
	\centering
	\includegraphics[width=0.8\textwidth]{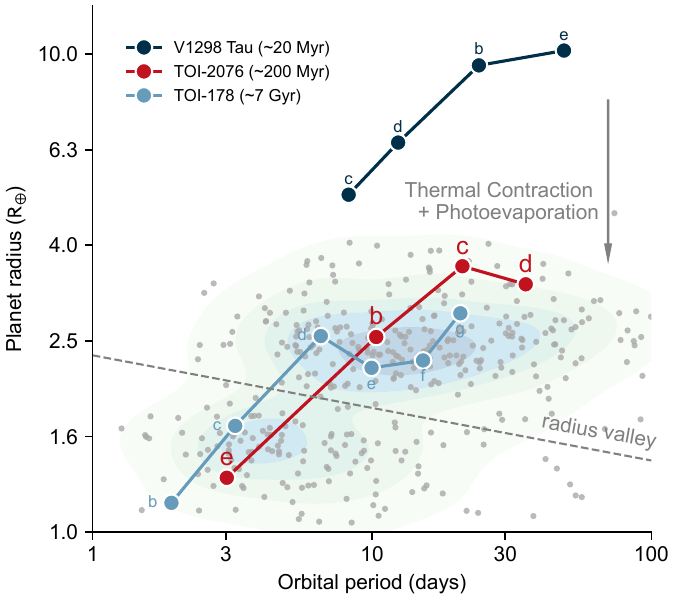} 
	\caption{\textbf{Comparison of planetary radii and orbital periods for V1298 Tau, TOI-2076, and TOI-178.}
    Large coloured circles show the planets in the three systems, with letters indicating individual planets. Small gray points represent the \emph{Kepler} planets from California-Keck Survey, with the gray dashed line marking the location of radius valley. The downward arrow shows the direction of planetary radius evolution with age in the first hundreds of million years due to combined effects of photoevaporative mass-loss and thermal contraction.}
	\label{fig:sub_neptune_evol}
\end{figure}

We compare TOI-2076 to the younger near-resonant system V1298 Tau \cite{David2019,Livingston2026} and the more evolved resonant chain TOI-178 \cite{Leleu2021} in Figure~\ref{fig:sub_neptune_evol}, both of which likely have not experienced giant impacts. Young planets such as V1298 Tau are inflated due to their high internal entropy and massive primordial envelopes. Over the first several hundreds of Myr, cooling and atmospheric mass loss reduce their sizes, yielding adolescent sub-Neptune systems like TOI-2076. A similar trend is observed at the population level, where super-Neptune–size planets become increasingly rare within the first $\sim$100s Myr \cite{Vach2024,Fernandes2025,Dai2025UPiCII}. Thereafter, radius evolution slows down, and planetary radii remain nearly constant over Gyr timescales, as observed in mature systems such as TOI-178. The $\sim$200-Myr-old TOI-2076 hosts both super-Earths and sub-Neptunes (with H/He envelope mass fractions of order $\sim$1\%), suggesting that the bimodal radius distribution observed in mature systems \cite{Fulton} is already emerging by this age.

%%%%%%%%%%%%%%%% MAIN TEXT FIGURES %%%%%%%%%%%%%%%

\newpage
\bigskip

\begin{table}[!h]
\fontsize{10pt}{10pt}\selectfont
\centering
\caption{\textbf{Physical Parameters of TOI-2076 Planets}}
\label{tab:derived_physical_parameter}
\begin{tabular}{lllll}
\hline
\hline
Parameter & Planet e &  Planet b & Planet c & Planet d \\
\hline
$M_p$ ($M_\oplus$)	              & 	 $4.7 \pm {1.5}$  			& 	 $6.7 \pm {1.8}$  			& 	 $7.2 \pm {1.4}$  			& 	 $7.3 \pm {2.7}$  			\\ 
$R_p$ ($R_\oplus$)	              & 	 $1.301 \pm {0.059}$  		& 	 $2.59 \pm {0.1}$  			& 	 $3.54 \pm {0.14}$  		& 	 $3.27 \pm {0.13}$  	    \\
$\rho_p$ (g/cm$^{3}$)             & 	 $11.5 \pm {3.7}$  			& 	 $2.08 \pm {0.58}$  		& 	 $0.87 \pm {0.18}$  		& 	 $1.12 \pm {0.43}$  		\\ 
$a_p$ (AU)			              & 	 $0.03863(36)$  			& 	 $0.08781(82)$  			& 	 $0.1407(13)$  				& 	 $0.1982(19)$  				\\ 
$P_{\rm orb}$ (days)	          & 	 $3.0223753(79)$  			& 	 $10.35504(14)$			  	& 	 $21.01447(60)$  			& 	 $35.12803(59)$  			\\ 
$e_p$			                  & 	 $\equiv0$  				& 	 $0.0114 \pm {0.0054}$  	& 	 $<0.036$  					& 	 $<0.027$  					\\ 
$i_p$ (deg)		                  & 	 $89.46 \pm {0.42}$  		& 	 $89.75 \pm {0.18}$  		& 	 $89.76 \pm {0.12}$  		& 	 $89.164 \pm {0.014}$  		\\ 
$T_{\rm eq}$ (K)	              &	     $1110 \pm 24$              & 	 $736 \pm 16$  			    & 	 $581 \pm 13$  			    & 	 $490 \pm 11$  	            \\  
\hline
\multicolumn{5}{p{0.95\textwidth}}{
Orbital parameters are osculating parameters defined in BJD=2458743. Uncertainties denote the 68\% credible interval and upper limits are 2$\sigma$. Equilibrium temperatures are calculated assuming zero albedo. Planet radii are derived by convolving the radius ratio with $R_*=0.758\pm0.030 ~R_\odot$, the uncertainty is inflated to account for systematic uncertainty of 4\% between stellar models \cite{Tayar2022}.
}
\end{tabular}
\end{table}

\newpage

\setcounter{figure}{0}
\renewcommand{\figurename}{Extended Data Figure}

\begin{figure}[!ht]
    \hspace{-0.5cm}
    \includegraphics[width=1.0\columnwidth]{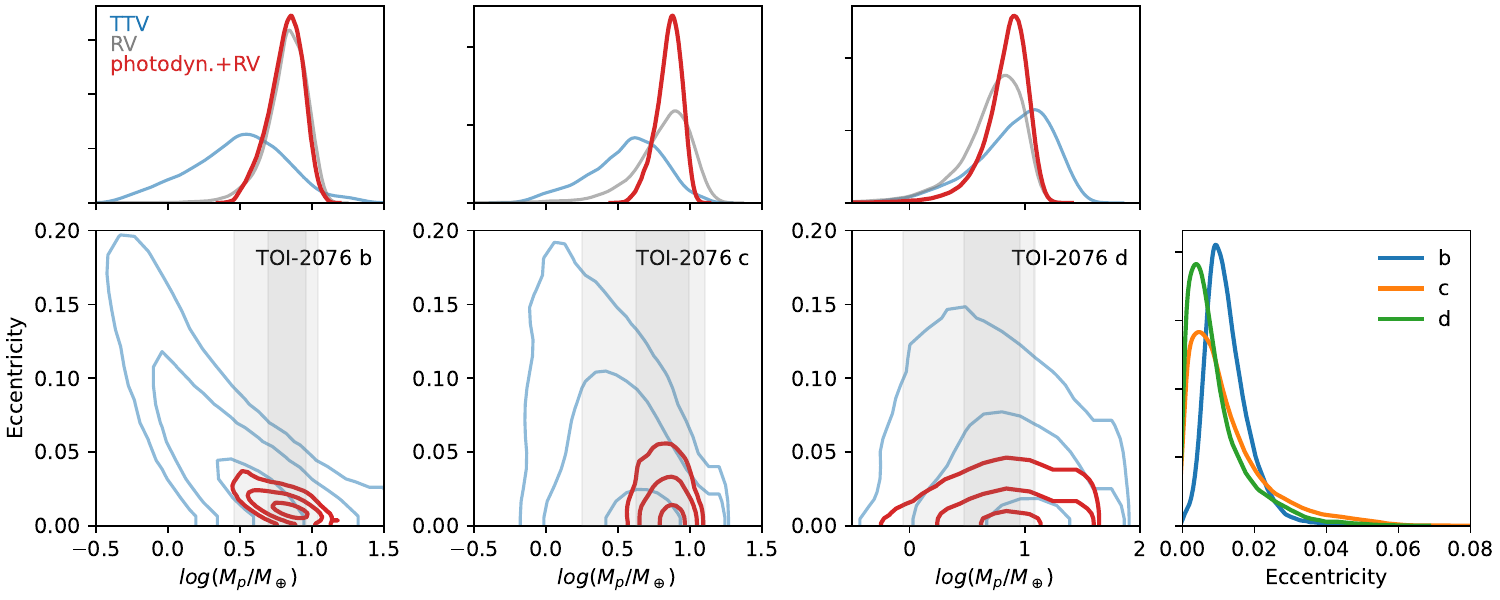}
    \caption{\textbf{Planetary mass and eccentricity constraints from different observational sources.} Top: marginalized planet mass constraints from TTV (blue), RV (gray), and joint photodynamical and RV model (red). First three panels in bottom show the planet mass and eccentricity 1 and 2$\sigma$ contours plot, color indexed as in top panels. Bottom rightmost panel shows the marginalized distribution of eccentricity from the joint photodynamical and RV model.}
    \label{fig:ttv_rv_photodyn_compare}
\end{figure}

\begin{figure}[!ht]
	\includegraphics[width=1.0\textwidth]{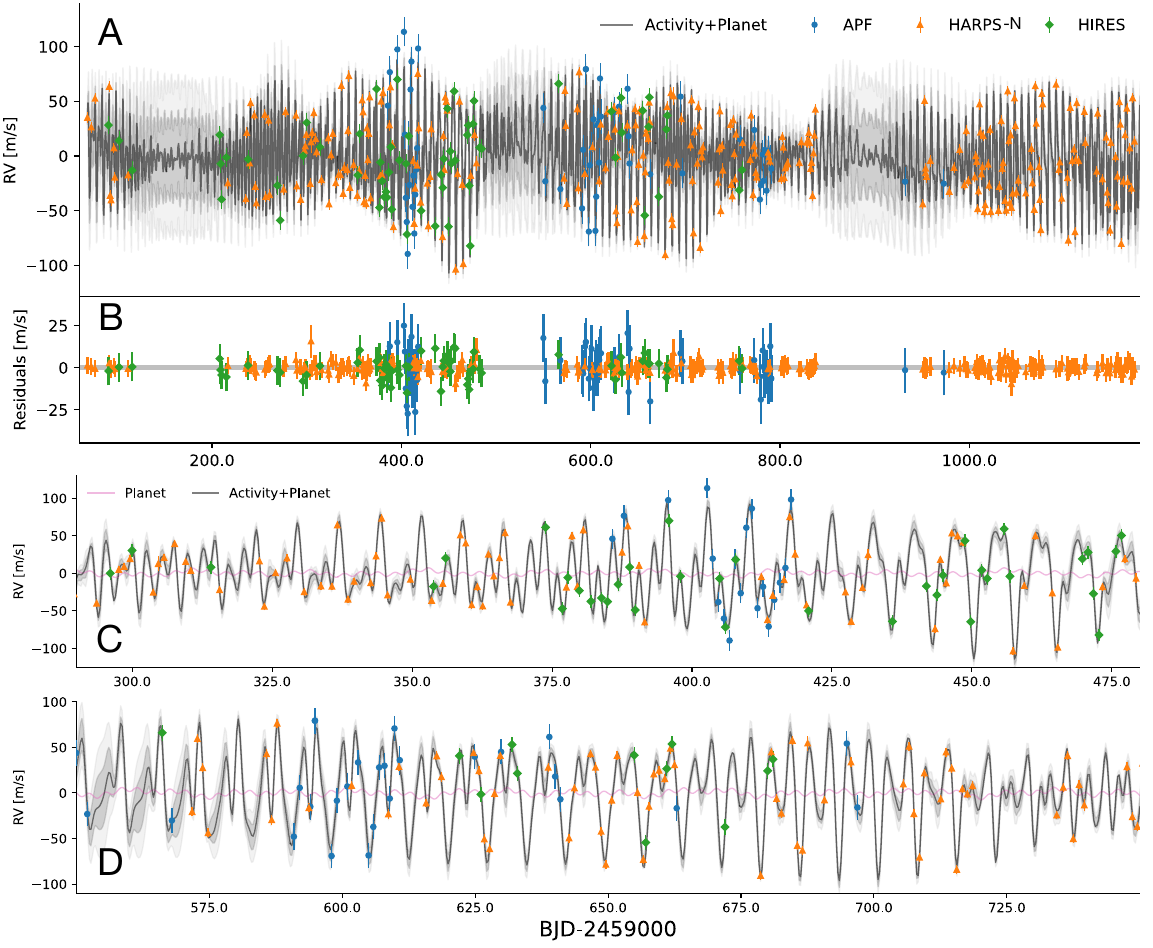}
        \hspace{-1.5cm}
	\caption{\textbf{Radial velocity data modeled with Keplerian planet signal and stellar activity with Gaussian Process.}
        (\textbf{A}) An overview of RV analysis of TOI-2076 with a Gaussian process (GP) activity model over the combined HIRES (orange triangle), APF (blue circle), and HARPS-N (green diamond) RV series of TOI-2076, overlaid with mean prediction from GP activity
        model (black line). The dark and light gray region show the 1$\sigma$ and 2$\sigma$ uncertainty band of GP model.
        (\textbf{B})RV residuals after subtracting the activity and planetary signals. Legends are the same as those in panel A.
        (\textbf{C-D}) zoom-in view of GP model at two observing seasons. Legends are the same as those in panel A.
    }
	\label{fig:rv_gp} % give each figure a logical label name
\end{figure}

\newpage
\section{Methods}\label{sec11}

\subsection{Photometric Observation}

\subsubsection{TESS Observation}
TESS observed TOI-2076 in Sector 16, 23, 50, and 77.  We downloaded the 20-second (sectors 16 and 23) and 2-minute (sectors 50 and 77) cadence light curve produced by the TESS Science Processing Operations Center (SPOC, of the NASA Ames Research Center; \cite{Jenkins2016}). The data are available on the Mikulski Archive for Space Telescopes website. Our analysis made use of the Simple Aperture Photometry (SAP \cite{SAP_Twickens_2010}) light curve. We applied systematic corrections following reference \cite{Vanderburg2019}. This method has shown to improve the noise levels in the light curves of young stars, enabling the recovery of previously missed transits (e.g. \cite{Barber}), and was the extraction method used in the discovery of TOI 2076 e \cite{Barber2025}. 

\subsubsection{LCO Observation}
The \textit{TESS} Follow-up Observing Program Sub Group 1 \cite{collins:2019} observed 11 full and partial transit windows of all three planets in Pan-STARRS $z_s$ band from the Las Cumbres Observatory Global Telescope (LCOGT) \cite{Brown:2013} 1m network nodes at McDonald Observatory near Fort Davis, Texas, United States (McD) and Teide Observatory on the island of Tenerife (TEID). We used the {\tt TESS Transit Finder}, which is a customized version of the {\tt Tapir} software package \cite{Jensen:2013}, to schedule our transit observations. The 1\,m telescopes are equipped with $4096\times4096$ SINISTRO cameras having an image scale of $0.389^{\prime\prime}$ per pixel, resulting in a $26^\prime \times 26^\prime$ field of view. We observed a partial transit window of TOI-2076d simultaneously in Sloan $g'$, $r'$, $i'$, and Pan-STARRS $z$-short from the Las Cumbres Observatory Global Telescope (LCOGT) \cite{Brown:2013} 2\,m Faulkes Telescope North at Haleakala Observatory on Maui, Hawai'i. The telescope is equipped with the MuSCAT3 multi-band imager \cite{Narita:2020}. All images were calibrated by the standard LCOGT {\tt BANZAI} pipeline \cite{McCully:2018} and differential photometric data were extracted using {\tt AstroImageJ} \cite{Collins:2017}. In all observations, we used circular photometric apertures that excluded all of the flux from the nearest known neighbor in the Gaia DR3 catalog (Gaia DR3 1490842629443812096), which is $\sim25^{\prime\prime}$ southwest of TOI-2076. All lightcurve follow-up observations are tabulated in Supplementary Table 1.

\subsubsection{CHEOPS Observation}
TOI-2076 was observed by CHaracterising ExOPlanet Satellite (CHEOPS, \cite{CHEOPS}) with 4 visits between 2021-2-28 to 2022-5-11, two of which are published in \cite{Osborn2022}. We downloaded the CHEOPS data from online archive. The raw data of each visit were automatically processed with the CHEOPS Data Reduction Pipeline (DRP, \cite{CHEOPS_DRP_2020}). 

\subsection{Light Curve Detrending and Transit Midtime Extraction}

The TESS light curves are trimmed to discrete segments centered around each individual transits with a width of three times the transit durations. We used 2-minute cadence data for all sectors and removed data with bad quality flags (\texttt{quality} $>$ 16). TESS Sector 77 observation was truncated due to on-board data recorder fill, Safe Hold mode, and data downloads. Only one transit of TOI-2076 b has been collected. 
Each segment of light curve has a local linear correction to account for out-of-transit variability. Robust detrending usually requires precise knowledge of the in-transit part of the light curve to avoid distortion on the transit shape. Therefore, we performed detrending in an iterative process. First, we fit a limb-darkening transit model and a linear slope simultaneously, with in-transit data given zero weight during the fit. We divide the light curve by the fitted slope to get the normalized data. Then, when our preliminary photodynamical model produces a good fit to the light curves, we used the more precise in-transit boundary informed by the dynamical model to perform the local detrend again, and the photometric uncertainties are rescaled to let the reduced $\chi^2$ of the light curve residuals equal one.

Instrumental systematics affect LCO and CHEOPS light curves more than TESS light curves, and a simple local linear correction against time is often insufficient. Instead, effective detrending requires decorrelation against one or more systemic parameters. We describe our decorrelation strategy separately.

LCO light curves are modeled with a Mandel \& Agol transit model \cite{Mandel2002} as implemented in  \texttt{batman} \cite{Kreidberg2015}, with free parameters including transit midtime $T_{\rm mid}$, star-planet radius ratio $p$, quadratic limb darkening coefficients, semi-major axis normalized to stellar radius ($a/R_*$), and additional photometric jitters. $a/R_*$ and $p$ are given Gaussian priors informed by transit modeling from TESS light curves. The systematic variations in light curves are detrended using a linear combination of ancillary parameters, e.g., observation time, airmass, total counts, and the target's pixel location. The raw and detrended LCO light curves are show in Supplementary Figure 1. The detrending parameters adopted for each LCO light curve are summarized in Supplementary Table 1.

The detrending of the CHEOPS light curves follows a similar approach to that of the LCO light curves. The trend induced by systematics is modeled with additive factors multiplied with the background flux, smears, and spacecraft roll angles. We used \texttt{lmfit} to find the factors and transit parameters best fitted for the CHEOPS light curves. The final best-fit model, incorporating both systematic trends and the transit model, along with the corresponding detrended light curves, is shown in Supplementary Figure 2.

We used the python package \texttt{juliet} \cite{Espinoza2019} to determine the transit times. The transit light curves are modeled with planet radius ratio, impact parameters, periods, and stellar density. Each transit time is treated as an independent fitting parameter but shares the same stellar density, which is given a Gaussian prior of 2.70 $\pm$ 0.15 g/cm$^3$. Photometric jitters and quadratic limb darkening coefficients are also assigned to each instrument. The posteriors of transit midtimes are obtained by \texttt{multinest}, and listed in Supplementary Table 2.

\subsection{Radial Velocity Analysis \label{sec:rv_analysis}}

We collected the archival RV data from HARPS-N, HIRES, and APF from \cite{Damasso2024,Polanski2024}.

For the RV analysis, we modeled the planetary Keplerian signals and stellar activity simultaneously. The stellar activities are modeled with Gaussian Process regression (GP), for which we employ the quasi-periodic kernel \cite{Grunblatt2015}:
\begin{align}
    \label{eq:gp_cov}
    C_{i,j} = & h^2  \exp\left [ -\frac{(t_i-t_j)^2}{2\tau^2} - \Gamma \sin^2\frac{\pi(t_i-t_j)}{T} \right ] \\ \nonumber
    &+ [\sigma^2_i+\sigma_{\rm jit}^2]\delta_{i,j} 
\end{align}
where $C_{i,j}$ is the element of covariance matrix, $\delta_{i,j}$ is the Kronecker delta function, $h$ is the covariance amplitude, $t_i$ is the time of $i$-th RV measurement, $\tau$ is the characteristic lifetime of active region, $\Gamma$ quantifies the relative importance between the exponential decay and periodic parts of the kernel, and $T$ is the stellar rotation period. We also introduce instrument-specific RV jitter terms $\sigma_{\rm jit}$ to account for additional noise.

The likelihood function of GP regression is:
\begin{equation}
\label{eq:gp_lh}
	\log \mathcal{L_{\rm RV}} = -\frac{N}{2}\log2\pi - \frac{1}{2}\log|\boldsymbol{C}| - \frac{1}{2}\boldsymbol{r}^T\boldsymbol{C}^{-1}\boldsymbol{r}
\end{equation}
where $\mathcal{L}$ is the likelihood, $N$ is the number of data points, $\boldsymbol{C}$ is the covariance matrix, and $\boldsymbol{r}$ is the residual vector (the observed RV minus the Keplerian signal).

The variability observed in both RVs and photometry is closely connected, as stellar spots— the surface brightness inhomogeneities that modulate the observed flux and distort the RV signal—rotate in and out of view, inducing correlated variations in the two datasets \citep{Haywood,Aigrain2012,Grunblatt2015,Dai2019}. Therefore, we first trained the GP hyperparameters on TESS light curves. The parameter set is the same as those involved in Eq. \ref{eq:gp_cov}, for which we all applied log-uniform priors. Three sectors of light curves were used to train the GP parameters. We used \texttt{emcee} to sample the posterior of GP parameters, yielding  $T=7.34 \pm 0.04$ days. This value is used as prior for rotation period in the following RV analysis, and the rest of the hyperparameters ($\tau$, $\Gamma$, $h$) are left as free parameters. 

We modeled the RV signal due to four planets with circular Keplerians and stellar activities with the quasi-periodic kernel GP. We gave Gaussian priors to each planet's period and date of conjunction, informed by the transit fitting. To calculate the significance of the presence of planet RV signals in the RV data, the semi-amplitudes of each signal are allowed to be negative. Stellar activity signals are assumed to be the same across HIRES, APF, and HARPS-N data, with $h_{\rm RV}$ measuring the amplitude of activity-induced RV variation. The priors and posteriors of fitted parameters are summarized in Supplementary Table 3. 

The RV-only analysis on the combined dataset of three instruments provides reasonable constraints ($\gtrsim3\sigma$) on the mass of planets e and b for $4.89\pm 1.34$ M$_\oplus$ and $7.1\pm 1.9$ M$_\oplus$, respectively. However, the mass of the outer planet c and d is only marginally constrained to be $6.91\pm2.43$ M$_\oplus$ (2.58$\sigma$) and $6.11\pm2.65$M$_\oplus$ (2.12$\sigma$), respectively. 

For the nominal analysis here, the HARPS-N data were mainly derived from \texttt{SERVAL} \cite{SERVAL}  pipeline. As the HARPS-N data take the largest fraction (294/412) in the whole dataset, we experimented with HARPS-N RV extracted by other pipelines: DRS and line-by-line \cite{HARPS_lbl} analysis, both provided in \cite{Damasso2024}. The mass constraints from RV product derived from the these two independent pipelines agree with the values in Supplementary Table 3 within 1$\sigma$.

To further quantify the statistical significance for planetary signals in the RV data, we performed nested sampling analyses with \texttt{dynesty} \cite{dynesty2020} to calculate the marginalized log-likelihood (evidence $\log Z$) of a pure GP model against models with additional Keplerian components. Relative to the pure GP model, the GP + planet b/e and GP + all planet models increase the evidence by $\Delta\log Z \sim 1.1$ and $2.0$, respectively. These results indicate positive support for the inclusion of Keplerian signals in the RV model, though the strength of the evidence remains moderate. Nevertheless, the existence of transits for all four planets strongly motivates the presence of corresponding RV signals at the modeled ephemerides. We therefore adopt the GP + all-planet model as our fiducial description of the RV data, while acknowledging the limited statistical significance.

\subsubsection{Cross Validation}

Gaussian processes (GPs) are designed to capture correlated stellar activity signals across datasets. When a GP successfully models the underlying physical process, it should yield accurate predictions for contemporaneous data. Conversely, if the GP primarily fits stochastic noise, its predictive power diminishes—an issue known as overfitting \cite{Blunt}. In this section, we evaluate whether, and to what extent, the adopted GP model provides an accurate description of the underlying signal.

We adopt the model parameters corresponding to the median values in Supplementary Table 3 as a representative case. Following the strategy of \cite{Blunt}, we condition the GP on the HARPS-N dataset and then use it to predict the unseen, contemporaneous HIRES and APF data. The residuals between the GP mean predictions and the observations (in units of RV and GP uncertainties) are shown in Supplementary Figure 5. As expected, the conditioned HARPS-N data display a narrow dispersion around the predicted values. The HIRES/APF residuals show a broader distribution, indicating a modest degree of overfitting. Nevertheless, they remain centered near zero, suggesting that the GP still captures the correlated noise effectively. By contrast, severe overfitting would manifest as dispersed, nearly uniform residuals and a loss of predictive power on contemporaneous data, as demonstrated in \cite{Blunt}. We therefore conclude that the adopted GP model does not suffer from significant overfitting, and the inferred parameters of interest (e.g., planet mass) remain reliable and unbiased.

\subsubsection{Injection-Recovery Test} 
We performed an injection–recovery test of planetary RV signals by generating synthetic datasets that incorporate the stellar activity signal, observational and instrumental white noise, and the injected planetary signal. For the stellar activity, we adopted the mean prediction of the GP model conditioned on the observed data using the median solution listed in Supplementary Table 3. Next, we added circular planetary signals at the orbital period and conjunction epoch of each of the four planets, with semi-amplitudes $K=$ 0, 0.5, 1.0, 1.5, 2.0, 2.5, 3.0, and 4.0 m/s. Finally, we included Gaussian white noise with variance $(\sigma_{\rm rv}^2 + \sigma_{\rm jitter}^2)^{1/2}$. In each experiment, only one planetary signal was injected at a time, and we conducted four realizations for each injected $K$ for each planet. The resulting synthetic datasets were then refitted following the same procedure outlined in Section \ref{sec:rv_analysis}.

The median and 1$\sigma$ uncertainties of the recovered planetary semi-amplitudes $K$ are plotted against the injected K in Supplementary Figure 4. The results show that the recovered values are scattered around 1:1 line, and are on average within 1$\sigma$ consistent with the injected values for planet e, b, c, and d. Therefore, we conclude that the recovered semi-amplitudes show no significant systematic bias (average offsets $\lesssim1\sigma$), and consider the reported semi-amplitudes in Supplementary Table 3 to be robust within this range.

\subsection{Photodynamical Modeling}

To disentangle and assess the planetary mass information encoded in the TTV data alone, we first perform a TTV-only retrieval using a simplified dynamical model. This allows us to evaluate the extent to which the observed timing variations independently constrain planetary masses, and to compare these results with those obtained from radial velocity (RV) measurements. We then carry out a joint modeling that incorporates both TTV and RV data to obtain refined constraints on planetary masses and eccentricities by leveraging the full set of observational evidence.

\subsubsection{Transit Timing Variations} The TTV model is initialized by the osculating orbital parameters defined in BJD = 2458743, including orbital period $P_i$, eccentricity vectors ($e\sin\omega,e\cos\omega$), and time of transit conjunctions $T_{\rm conj}$. As the TTV observation is sensitive to the planetary mass ratios rather than the absolute mass, we directly sampled planet-to-star mass ratios, and revert back to planet mass after obtaining the posterior distribution. We assume coplanar orbits, so the ascending node angles $\Omega$ are fixed at zero and inclination $i$ at $\pi/2$. 

We did not include TOI-2076 e in the TTV-only model, as it does not display notable TTV signals in TESS observations. It is also dynamically detached from planet b, c, and d because it is not in or close to any strong MMRs with them ($P_b/P_e=3.42,~P_c/P_e=6.95,~P_e/P_e=11.62$), therefore it will not significantly contribute to the observed TTVs, which are mainly due to adjacent planet interactions. 

We used \texttt{jnkepler} to calculate the model TTVs, which performs N-body integration using a symplectic integrator \cite{Masuda2024}. The resulting orbits are used to derive mid-transit times of each planet where the planet-star distance in the sky is minimized. We use a time step of 0.3 days for the symplectic integrator, the resulting timing precision is $\sim$ 1 second. 
    
All parameters are assigned with a uniform prior. The planet orbital eccentricities $e$ are assigned with a prior probability $\sim1/e$ to cancel the explicit prior of $\sim e$ when uniformly sampling the eccentricity vectors $e\cos\omega$ and $e\sin\omega$.  
    
The likelihood function is defined as chi-squared-like:
    \begin{equation}
        \log \mathcal{L_{\rm TTV }} = -\frac{1}{2}\sum_{i} \left ( \frac{t_{\rm i,obs}-t_{\rm i,model}}{\sigma_{\rm i,obs}} \right )^2
    \end{equation}
where the $t_{\rm i,obs}$ and $\sigma_{\rm i,obs}$ are the observed transit midtimes and uncertainties listed in Supplementary Table 2, $t_{\rm i,model}$ is the transit midtimes calculated by TTV model.

We used the No-U-Turn sampler implemented in \texttt{numpyro} \cite{numpyro2019} to sample the posterior. We ran 64 chains with 500 burn-in steps and 1500 sampling steps. The Gelman-Rubin statistics $\hat{R}$ \cite{Brooks1998} for all parameters is smaller than 1.01, and the effective sample size exceeds 1000, indicating good convergence. The priors, medians and 68\% credible intervals are summarized in Supplementary Table 4.

We found that the TTVs alone yielded a wide range of planetary mass and eccentricity due to the inherent degeneracy of near-resonant dynamics. In this case, only the product of mass and eccentricity is constrained, while their individual values can vary significantly \citep{Lithwick_ttv,Hadden2017,Leleu2023}. To assess the robustness of TTV-derived mass and eccentricity, we tested three sets of priors: (1) the default prior, which assumes uniform distributions for both mass and eccentricity (from direct sampling); (2) the high-mass prior, which assumes a uniform prior for mass and a log-uniform prior for eccentricity; and (3) the low-mass prior, which assumes a log-uniform prior for mass and a uniform prior for eccentricity. The samples corresponding to the latter two priors were obtained by resampling the first set of samples, with each sample weighted by 1/$m$ or 1/$e$. Comparisons of the three sets of samples are shown in Supplementary Figure 5. We observe that different priors yield statistically distinct planetary parameters, with differences exceeding $1\sigma$ between the low-mass and high-mass priors. Therefore, the planetary mass and eccentricity derived from TTV data alone are highly degenerate and sensitive to the adopted priors. Joint photodynamics modeling that incorporates the infomation in TTV, RV, and photoeccentric effect might help to break the degeneracy.

\subsubsection{Photodynamical Modeling} We modeled the available photometry data and radial velocity data of TOI-2076 using a photodynamical model, which include a dynamical model that accounts for planetary mutual gravitational interactions and predicts planetary transit light curves. Our model follows the framework of \cite{Wang2024}, as detailed below.

Each planet orbit is initialized with osculating parameters of orbital period $P$, eccentricity $e$, argument of pericenter $\omega$, impact parameter $b$, and time of inferior conjunction $T_c$. The eccentricity and argument of the pericenter are parameterized as $e\cos\omega$ and $e\sin\omega$ to improve the sampling efficiency due to their inherent correlation from near-resonant TTVs \cite{JontofHutter2016}. The ascending nodes $\Omega$ are all set to 0. The system is described in Jacobian coordinates, with osculating orbital parameters referenced to BJD = 2458743.0. We integrate the system’s motion using the symplectic \texttt{WHFast} integrator \cite{whfast2015} within the \texttt{rebound} package \cite{ReinLiu2012}, considering only Newtonian gravitational interactions. The integration timestep is fixed at 0.05 days, approximately 1/60 of the innermost planet’s orbital period, and is performed in the center-of-mass frame. The $+z$-axis is defined as pointing toward the observer. At each observed epoch of photometric and radial velocity data, we record the barycentric coordinates and velocities of all bodies, using the z-component of stellar velocity to compare with radial velocity measurements. 

Light curves are synthesized from the relative coordinates of different bodies with the Mandel \& Agol quadratic limb-darkening model \cite{Mandel2002}, implemented in \texttt{batman} \cite{Kreidberg2015}. We ignored light travel time effects (LTTEs) as its correction on transit timing ($\lesssim$1.6 minute) are smaller than the observational uncertainty. The total loss of light is assumed to be the sum of losses due to all planets, without accounting for planet-planet occultations \cite{Masuda2013}. Four planetary-to-star radius ratios ($R_{\rm e, b,c,d}/R_s$) and three sets of quadratic limb darkening coefficients for TESS, CHEOPS, and LCO $z_s$ bands are assigned for light curve synthesis.  

Finally, we adopted a Gaussian prior of 0.758$\pm$0.014 R$_\odot$ for stellar radius, and 0.849$\pm$0.026 M$_\odot$ for stellar mass, both of which are determined by \cite{Damasso2024}. The planet orbital eccentricities $e$ are assigned with a prior probability $\sim1/e$ to cancel the explicit prior of $\sim e$ when uniformly sampling the eccentricity vectors ($e\cos\omega$, $e\sin\omega$). The prior bounds are summarized in Supplementary Table 4.

The log-likelihood of photometric data, $\mathcal{L_{\rm pho}}$, follows a $\chi^2$ distribution:

\begin{equation}
\label{eq:photo_lh}
    \log \mathcal{L_{\rm pho}} = -\frac{1}{2}\sum_{i}^{N} \left ( \frac{f_{\rm i,obs}-f_{\rm i,model}}{\sigma_i} \right )^2
\end{equation}
where $f_{\rm i,obs}$ is the observed normalized fluxes, $f_{\rm i,model}$ is the fluxes predicted by the dynamical model, $\sigma_i$ is the flux error, and $N$ is the total number of photometry data. 

For the joint photodynamical and RV joint modeling, ten more parameters, as listed in Supplementary Table 3, are required, thus there are in total 38 adjustable parameters  for the joint model. The likelihood of the joint model is the linear combination of Eq \ref{eq:photo_lh} and Eq \ref{eq:gp_lh}. 

To identify potentially different modes of posteriors, we perform dynamic nested sampling with \texttt{dynesty} \cite{dynesty2020}. We adopt the multiple-ellipsoid method to update iso-likelihood bounds. We initialize 2000 live points and perform two independent runs until the difference of evidence between iterations is smaller than 10$^{-4}$. The two sets of posteriors for the above run have consistent posterior distributions and are far from prior bounds. 
The resulting posterior was used to initialize the walkers for an MCMC sampling routine implemented in \texttt{emcee}. The MCMC is initialized with 100 walkers for another 500000 steps \cite{emcee}. We discard the first 20\% of chains as burin-in and thin down the samples every 2000 steps. The Gelman-Rubin statistics for all parameter are $<1.03$ and 700-4800 effective samples for different parameters \cite{Brooks1998}. The light curves model with best-fit solution is shown in Supplementary Figure 6. The median and 68\% intervals of the posterior are reported in Supplementary Table 4 and 5. The joint and marginalized posterior distributions of fitted stellar and planetary parameters are shown in the Supplementary Figure 7. The mass and eccentricity posterior distributions are shown in Extended Data Figure 1.

\subsection{Initial H/He Envelope Contents}
We discuss the initial envelope mass for TOI-2076 planets in the framework of both dusty and dust-free accretion. If the primordially accreted atmosphere is dusty, the initial gas-to-core mass ratio (GCR) scales with $\Sigma_{\rm neb}^{0.12}~M_{\rm core}^{1.7}t^{0.4}$, with $\Sigma_{\rm neb}$ as local nebular gas density, $M_{\rm core}$ as the core mass, and $t$ as the accretion time \cite{Lee2019}. The initial GCR depend strongly on planet core mass, therefore it is expected the GCR of planet bcd should be similar, as their core mass do not vary much, while planet e could have started more gas-poor. Assuming accretion time of 5 Myr, the initial GCR for TOI-2076 e would be 2.5-4\%, and 6–10\% for TOI-2076 b,c,d. The estimated range is mainly attributed to the uncertainty in core mass.

If the accreted gas were dust-free—implying reduced envelope opacity due to dust sedimentation or growth—the initial GCR would further vary with equilibrium temperature as GCR $\propto$ $T_{\rm eq}^{-1.9}$ \cite{Lee2015}. This results in a $\times1.8$ times increase to the initial GCR of the coldest planet, TOI-2076 d, relative to TOI-2076 b. However, the precise initial and current atmospheric properties remain uncertain. Future JWST transmission spectroscopy could provide further constraints. 
For simplicity, we assume all planets begin with the same initial GCR but explore a range of 5–10\% to account for these uncertainties in the following atmospheric evolution simulations.

\subsection{Evolution of Planetary H/He Envelopes}

We use the MESA planets package updated by \cite{Gu+Chen23} (further modified from \cite{Malsky20} and \cite{Chen2016}) to simulate super-Earth sized planets with a few percent of mass in their H/He envelopes. Their model includes diffusion separation of hydrogen/helium/deuterium, but for the purposes of the study, we do not include composition fractionation processes and focus on bulk escape. Photon-limited and radiative-recombination-limited escape were also not considered. We adopt the energy-limited escape scheme to approximate the time changing mass loss rate of planetary atmospheres around chromospherically active hosts. High energy irradiation (including extreme ultraviolet and X-ray radiation (XUV; $1 \lesssim \lambda \lesssim 1200$ \r{A}) imparted on the atmosphere of the young planets contributes to heating and hydrodynamic expansion, leading to some gas escape its gravity well. 

The energy-limited mass loss rate is given by  \cite{Watson1981,ErkeavEt2014,Murray-Clay2009}:
\begin{equation} 
\frac{{\rm d} M_p}{{\rm d} t} = - \frac{\epsilon_{\rm XUV} \pi F_{\rm XUV}  R_{p}^3}{G M_p K_{\rm tidal}}
\end{equation}

\noindent where $\epsilon_{\rm XUV}$ is the mass loss efficiency (i.e. the fraction of incident XUV energy that contributes to unbinding the outer layers of the planet), which depends on envelope composition, the XUV flux, and the XUV energy spectra \cite{Jackson2012}.  $M_p$ is the total mass of the planet and $G$ is the gravitational constant. $R_p$, or the radius of the planet. $R_{p}$ and $M_p$ are planet radius at optical depth $\tau_{\rm visible} = 1$ (in the visible) and the total mass of the planet respectively. $G$ is the gravitational constant. $R_{\rm Hill} \approx a \left({M_p}/{3 M_{*}}\right)^{1/3}$ represents the distances out to which the planet's gravitational influence dominates over the gravitational influence of the star. 

In our calculations, we assume $R_{\rm Hill}$ to be located well within the exobase where the particle mean free path and atmospheric scale height are comparable. $K_{\rm tidal}$ corrects for tidal forces, which modify the geometry of the potential energy well and decrease the energy deposition needed to escape the planet's gravity. The specific dependencies of $K_{\rm tidal}$ on  of $R_{\rm Hill}$ and $R_{\rm XUV}$ can be found in \cite{ErkeavEt2014}.

Finally, $R_{\rm XUV}$ is the distance from the center of the planet to the point where the atmosphere is optically thick to XUV photons. $F_{\rm XUV}$ is the extreme ultraviolet energy flux from the host star impinging on the planet atmosphere, which we approximate via \cite{Ribas2005,Valencia2010}:

\begin{equation} 
F_{\rm XUV} = 29.7~ \rm erg~s^{-1}~cm^{-2} \left(\frac{t}{1\rm Gyr}\right)^{-1.23}\left(\frac{a}{1{\rm~AU}}\right)^{-2}
\end{equation}
\noindent where $t$ is the stellar age in Gyr, $a$ the planet's orbital distance in AU, and $\alpha$ and $\beta$ are constants that depends on stellar characteristics and spectral type. For stellar spectral types between G and K, we adopt $\alpha = $ and $\beta = 1.23$. 

All planets are assumed to have Earth-like silicate-iron cores and to be undergoing pure energy-limited escape, with disk clearing ages of ${\sim}1$ Myr. For all three planet scenarios, we use initial metallicities $Z = 0.03$ and helium fractions $Y = 0.25$. In order for the planets to have consistent initial starting conditions, we inflate each planet to their maximum allowable entropy and then allow their envelopes to cool for $10^5$ yrs before adding the stellar irradiation to each planet model. A similar prescription has also been used by \cite{Malsky20}  and \cite{Chen2016}. The orbital separations are 0.04, 0.088, 0.141, and 0.198 AU, respectively for planets e, b, c, and d.  We test a range of mass-loss efficiencies, between 0.05 and 0.2. We simulate a grid of planets between 5 and 9 M$_\oplus$ initial planet masses, envelope fractions of 5-10\%, and mass-loss efficiencies 0.05-0.2. 

Within this grid, we find that planet models initialized with envelope fractions of 6.5\% and the lowest mass-loss efficiencies (0.05) produced the best-fit to observations. Such low efficiencies suggest that there may be other processes, for instance, radiative recombination, that act to reduce the rate of escape at these high temperature regimes relative to the (potentially overestimated) values given by the energy-limited formalism. are in good agreement the observationally-derived radii and mass values. Specifically, our best-fit initial planet masses are 4.5, 7, 8.5, and 6 M$_\oplus$ respectively. Figure~\ref{fig:atm_evol} shows the evolution of radii and H/He fraction of this sample. With all three planets starting with the same H/He fraction of 6.5\%, their mass loss evolution tracks at this early stage ($< 500$ Myr) already display substantial differences. The mass-loss rate of TOI-2076 b from the simulation is $\sim 6\times10^{11} $ g/s, which is slightly larger but comparable to the order-of-magnitude mass-loss estimate from the helium absorption observation ($\sim 3\times10^{11}$ g/s) \cite{Zhang2023}.

We find that XUV-driven hydrodynamic escape offers a plausible explanation for the divergent atmospheric outcomes observed among the four modeled planets, even though they likely formed in broadly similar disk environments with comparable primordial compositions. However, this mechanism is unlikely to operate in isolation or to dominate in every case. Our models did not account for additional processes emphasized in previous studies, such as the enhanced irradiation during the pre-main-sequence phase \citep{tian2015} or internally driven mass-loss channels including boil-off and core-powered escape \citep{Owen_boiloff,Ginzburg}. These mechanisms are expected to further modify the evolution of planetary envelopes, and in combination with XUV-driven escape, likely shape the long-term thermal and compositional states of close-in planets \cite{Owen2024,rogers2024,Tang2025}. In addition, recent observations suggest a slower decay rate for the stellar XUV flux \cite{King2021}, implying that the integrated high-energy irradiation may be more strongly weighted toward gigayear timescales than previously assumed. This would mean that some sub-Neptunes could continue to lose substantial envelope mass and potentially evolve into stripped rocky cores at much later epochs. Finally, we did not consider the possibility of high-metallicity envelopes. Enhanced metal abundances can increase radiative cooling in the upper atmosphere, thereby reducing the efficiency of hydrodynamic escape and allowing planets to retain more of their primordial gas \cite{Tian2009}. Such scenarios could therefore alter the inferred atmospheric mass-loss histories compared to the solar-metallicity cases modeled here.

\section*{Data Availability} 
TESS data can be obtained from the Mikulski Archive for Space Telescopes: https://doi.org/10.17909/T9-NMC8-F686 (ref. \cite{MASTtesslc}). HARPS-N RV data can be accessed at https://doi.org/10.26093/cds/vizier.36900235 (ref. \cite{Damasso2024Vizier}). Keck and APF RV data, along with other data that can reproduce the figure in main text are archived in https://doi.org/10.6084/m9.figshare.29815580 (ref. \cite{Wang2025figshare}). Additional data products not included in these archives can be obtained from M. W. upon reasonable request.

\section*{Code Availability} This study makes use of the following publicly-available packages: \texttt{rebound}\citep{ReinLiu2012} (https://github.com/hannorein/rebound), 
\texttt{jnkepler}\cite{2025ascl.soft05006M} (https://github.com/kemasuda/jnkepler), 
\texttt{emcee}\cite{emcee} (https://github.com/dfm/emcee). 
MESA scripts for atmosphere evolution track will be provided upon reasonable request.

\section*{Acknowledgement}
We thank James Owen, Ruth Murray-Clay, Eric Agol, Yayaati Chachan, Kento Masuda, Heather Knuston, and Michael Zhang for helpful feedback and discussions on the manuscript. 
This work is supported by National Key R\&D Program of China, No. 2024YFA1611801 and science research grants from the China Manned Space Project with No. CMSCSST-2025-A16. M.W. acknowledge the support from National Natural Science Foundation of China (grant No. 124B2058).
M.G. acknowledges support from ERC grant No. 101019380 (HolyEarth).
This work has been carried out within the framework of the NCCR PlanetS supported by the Swiss National Science Foundation under grants 51NF40\_182901 and 51NF40\_205606. A.L. acknowledges additional support from the Swiss National Science Foundation under grant number TMSGI2\_211697. 
Y.A. acknowledges support from the Swiss National Science Foundation under grant 200020\_192038. S.G.S. acknowledges support from FCT through contract CEECIND/00826/2018 and POPH/FSE (EC); the Portuguese team also thanks the Portuguese Space Agency for financial support via the PRODEX Programme of the European Space Agency under contract 4000142255. 
A.B. was supported by the SNSA. P.M. acknowledges support from STFC research grant ST/R000638/1. M.G.B. and A.W.M. acknowledge support from NASA's exoplanet research program (80NSSC25K7148). M.G.B. also acknowledges support from the NSF Graduate Research Fellowship (DGE-2040435).
We acknowledge financial support from the Agencia Estatal de Investigaci\'on of the Ministerio de Ciencia e Innovaci\'on MCIN/AEI/10.13039/501100011033 and the ERDF “A way of making Europe” through project PID2021-125627OB-C32, and from the Centre of Excellence “Severo Ochoa” award to the Instituto de Astrofisica de Canarias.

\section*{Author Contribution} 
M.W. led the data analysis and performed the photodynamical simulations. 
F.D. supervised the project and facilitated dynamical modeling. 
H.C. conducted the planetary envelope evolution simulations. 
M.W., F.D., and H.C. wrote the manuscript. 
M.W., F.D., H.L., Z.H., and M.G. contributed to the interpretation of the dynamical results. 
E.A.P. contributed to the discussion on interpretation of the analytical TTV analyses. 
E.J.L. contributed to the discussion on the initial conditions of planetary envelopes. 
M.G.B. and A.W.M. assisted in TESS light curve extraction.
H.L., E.A.P., E.J.L., A.L., J.N.W., A.W.M., and S.G. contributed to scientific discussions and provided substantial feedbacks throughout the project.
K.A.C. coordinated LCO scheduling, contributed to data reduction, and provided observing time. C.N.W. coordinated SG1 observations, reduced LCO data, and write the manuscript. R.P.S., H.M.R., A.G., and F.P.W. performed LCO data reduction. E.P., F.M., R.S., and K.H. contributed LCO telescope time through their respective institutions. A.S. served as the PI of the LCO Key Project and oversaw coordination of observing resources. 
A.L., H.O., Y.A., L.F., A.F., S.S., A.B. and P.M. contributed to the scheduling of CHEOPS observation and data reduction.

\section*{Competing Interests}
The authors declare no competing interests.


\bigskip
\newpage

\section{Supplementary Information}

\setcounter{figure}{0}
\setcounter{table}{0}

\renewcommand{\figurename}{Supplementary Figue}
\renewcommand{\tablename}{Supplementary Table}

\begin{figure}[!h]
	\centering
    	\includegraphics[width=1.0\textwidth]{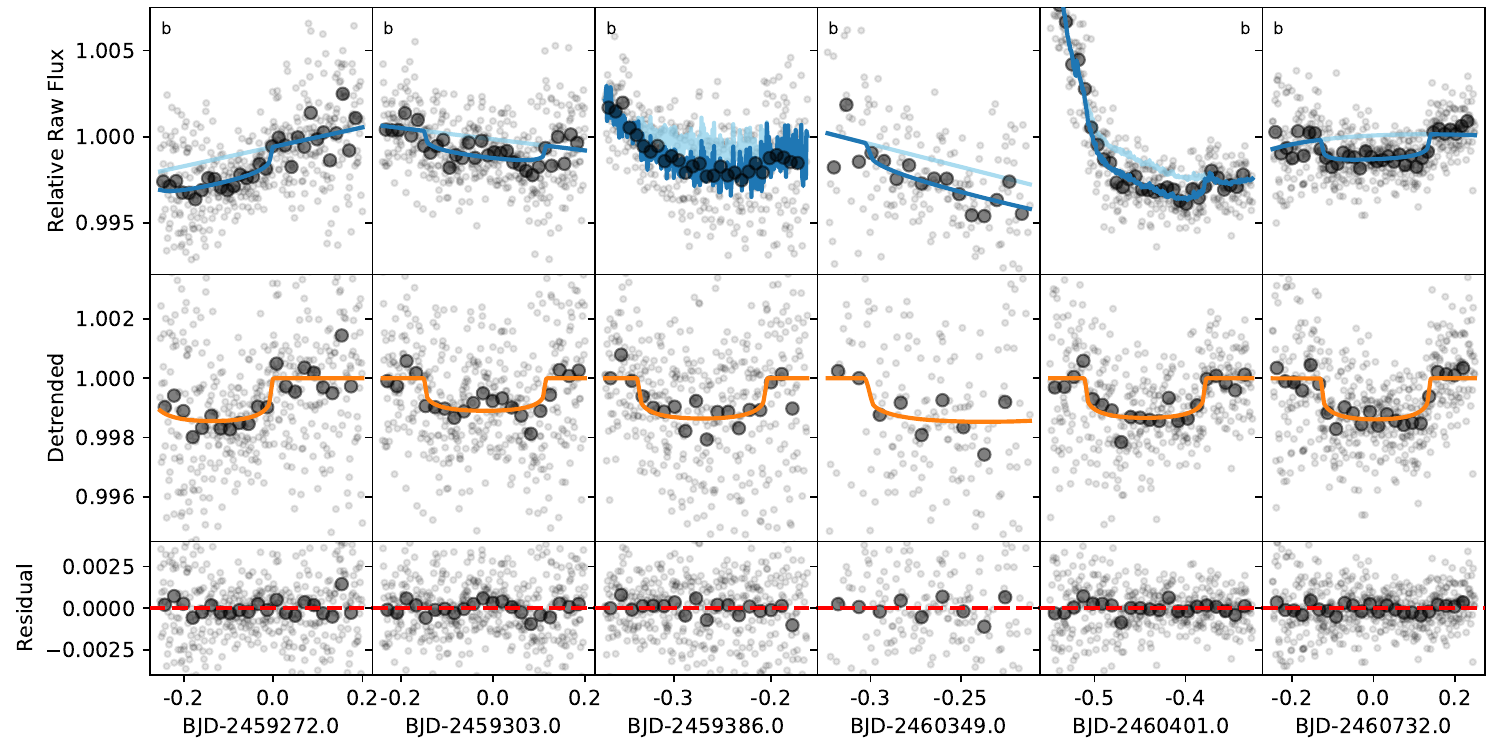} 
        \includegraphics[width=1.0\textwidth]{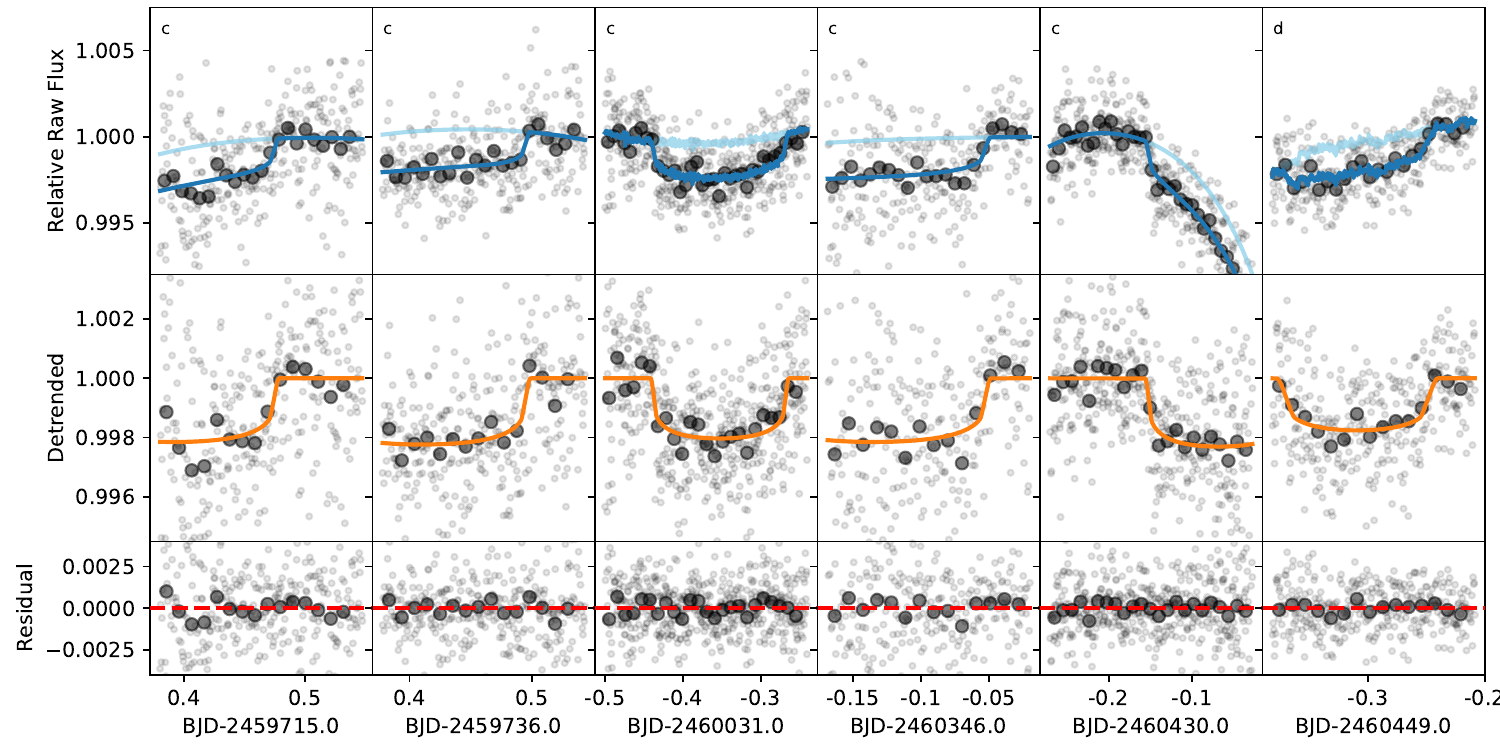} 
	\caption{\textbf{LCO light curves of TOI-2076 planets.}
        The LCO $z_s$ band light curves of TOI-2076 bcd planets. For each panel, the top row shows the raw fluxes (gray dots) and fluxes binned to 10 minutes (black dots), with the relevant planet index shown on top left. Blue lines are transit models with systematic-induced trends and light blue lines show the trends separately. Middle rows shows detrended light curves with best-fit transit model (orange), and bottom row shows light curve residuals.
    }
	\label{fig:LCO_lightcurve} % give each figure a logical label name
\end{figure}

\begin{figure}[!h]
	\centering
    	\includegraphics[width=1.0\textwidth]{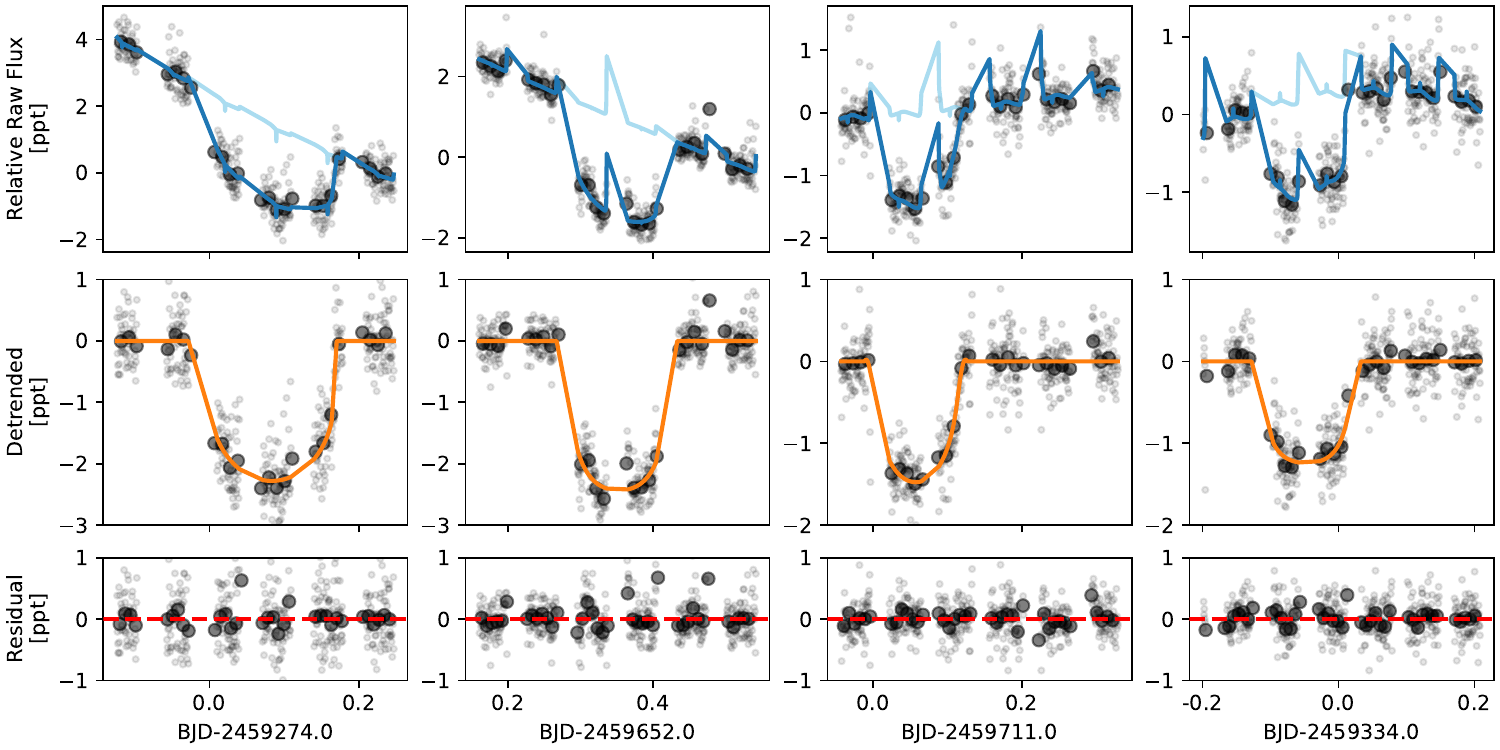} 
	\caption{\textbf{CHEOPS light curves of TOI-2076 planets.}
        Annotations are the same as Figure~\ref{fig:LCO_lightcurve}.
    }
	\label{fig:cheops_lightcurve}
\end{figure}

\begin{table}[h]
\centering
\caption{\textbf{Overview of LCO observations}.LCO-McD: McDonald Observatory, Texas, United States (McD); LCO-TEID: Teide Observatory, Tenerife (TEID)}
\label{tab:lco_log}
\begin{tabular*}{1.0\textwidth}{lcccccl}
\hline
\hline
Instrument & Start Date (UTC) & End Date (UTC) & Band & $N_{\rm Exp}$ & Planet & Type \\
\hline
LCO-McD   & 2021-02-26 06:33:53 &   2021-02-26 11:59:40  &   $z_s$     &   429     &   b    &    Full       \\  
LCO-McD   & 2021-03-29 06:42:47 &   2021-03-29 11:59:59  &   $z_s$     &   425     &   b    &    Full       \\
LCO-McD   & 2021-06-20 03:06:34 &   2021-06-20 08:05:49  &   $z_s$     &   392     &   b    &    Full       \\
LCO-Teid  & 2024-02-08 04:13:49 &   2024-02-08 06:55:12  &   $z_s$     &   176     &   b    &    Ingress    \\
LCO-Teid  & 2024-03-30 22:49:57 &   2024-03-31 04:09:38  &   $z_s$     &   405     &   b    &    Full       \\
LCO-McD   & 2025-02-25 06:37:09 &   2025-02-25 12:35:37  &   $z_s$     &   483     &   b    &    Full       \\
LCO-Teid  & 2022-05-15 21:07:50 &   2022-05-16 01:07:49  &   $z_s$     &   317     &   c    &    Egress     \\
LCO-Teid  & 2022-06-05 21:04:14 &   2022-06-06 01:01:47  &   $z_s$     &   322     &   c    &    Egress     \\
LCO-Teid  & 2023-03-27 00:00:10 &   2023-03-27 06:14:00  &   $z_s$     &   497     &   c    &    Full       \\
LCO-McD   & 2024-02-05 07:56:48 &   2024-02-05 11:31:15  &   $z_s$     &   270     &   c    &    Egress     \\
LCO-McD   & 2024-04-29 05:31:00 &   2024-04-29 11:19:53  &   $z_s$     &   435     &   c    &    Ingress    \\
LCO-McD   & 2024-05-18 02:50:21 &   2024-05-18 06:59:52  &   $z_s$     &   337     &   d    &    Full       \\
\hline
\end{tabular*} 

\end{table}

\begin{table}[h]
\centering
\caption{\textbf{Observed transit epochs and midtimes for all three planets in TOI-2076.} a: excluded from TTV analysis.  }
\fontsize{10pt}{10pt}\selectfont
\label{tab:obs_midtime}
\begin{tabular*}{\textwidth}{l@{\extracolsep{\fill}}ccl}
\hline
Epoch  & Transit Midtime                & Uncertainty         & Instrument \\
       & (BJD-2457000)                  & (days)              &            \\
\hline
\textbf{planet e} & & & \\
 0     & 1737.180700                    & 0.0032                & TESS         \\
 1     & 1740.203100                    & 0.0009                & TESS         \\
 2     & 1743.239800                    & 0.0044                & TESS         \\
 3     & 1746.257700                    & 0.0053                & TESS         \\
 4     & 1749.288600                    & 0.0041                & TESS         \\
 5     & 1752.297000                    & 0.0050                & TESS         \\
 6     & 1755.320600                    & 0.0046                & TESS         \\
 7     & 1758.357100                    & 0.0037                & TESS         \\
 63    & 1927.600300                    & 0.0075                & TESS         \\
 64    & 1930.619500                    & 0.0062                & TESS         \\
 65    & 1933.638800                    & 0.0065                & TESS         \\
 66    & 1936.667000                    & 0.0049                & TESS         \\
 68    & 1942.700100                    & 0.0019                & TESS         \\
 69    & 1945.738900                    & 0.0050                & TESS         \\
 70    & 1948.755000                    & 0.0044                & TESS         \\
 71    & 1951.787200                    & 0.0055                & TESS         \\
 308   & 2668.062800                    & 0.0024                & TESS         \\
 309   & 2671.096000                    & 0.0039                & TESS         \\
 310   & 2674.125300                    & 0.0047                & TESS         \\
 313   & 2683.174500                    & 0.0017                & TESS         \\
 314   & 2686.203600                    & 0.0052                & TESS         \\
 548   & 3393.439400                    & 0.0038                & TESS         \\
 549   & 3396.459100                    & 0.0057                & TESS         \\
 556   & 3417.610000                    & 0.0038                & TESS         \\
 557   & 3420.632000                    & 0.0100                & TESS         \\
\textbf{planet b} & & & \\
0      & 1743.7180                      & 0.0037              & TESS       \\
1      & 1754.0776                      & 0.0011              & TESS       \\
18     & 1930.1254                      & 0.0019              & TESS       \\
19     & 1940.4806                      & 0.0024              & TESS       \\
20     & 1950.8339                      & 0.0010              & TESS       \\
51     & 2271.8330                      & 0.0039              & LCO        \\
54     & 2302.8954                      & 0.0056              & LCO        \\
57     & 2333.9544                      & 0.0021              & CHEOPS     \\
62     & 2385.7322                      & 0.0045              & LCO        \\
89     & 2665.3387                      & 0.0016              & TESS       \\
90     & 2675.6944                      & 0.0013              & TESS       \\
91     & 2686.0499                      & 0.0012              & TESS       \\
155$^a$& 3348.7591                      & 0.0078              & LCO        \\
160    & 3400.5577                      & 0.0026              & LCO        \\
162    & 3421.2644                      & 0.0011              & TESS       \\
192	   & 3731.9025 	                    & 0.0018 	          & LCO        \\
\textbf{planet c} & & & \\
0      & 1748.69389                     & 0.00078             & TESS       \\
9      & 1937.82322                     & 0.00086             & TESS       \\
25     & 2274.08475                     & 0.00061             & CHEOPS     \\
43     & 2652.35118                     & 0.00049             & CHEOPS     \\
44     & 2673.36567                     & 0.00067             & TESS       \\
46     & 2715.39312                     & 0.00382             & LCO        \\
47     & 2736.41236                     & 0.00206             & LCO        \\
61     & 3030.64815                     & 0.00136             & LCO        \\
76     & 3345.86622                     & 0.00308             & LCO        \\
80     & 3429.92971                     & 0.00206             & LCO        \\
\textbf{planet d} & & & \\
0      & 1762.66666                     & 0.00173             & TESS       \\
5      & 1938.29185                     & 0.00218             & TESS       \\
18     & 2394.92676                     & 0.00120             & LCO        \\
26     & 2675.93174                     & 0.00115             & TESS       \\
27     & 2711.05693                     & 0.00082             & CHEOPS     \\
48     & 3448.69029                     & 0.00191             & LCO        \\
\hline
\end{tabular*}
\end{table}

\begin{table}[!h]
\centering
\fontsize{11pt}{11pt}\selectfont
\caption{\textbf{RV analysis Results with Gaussian Process.}
$\mathcal{N}(\mu,\sigma)$ is normal prior with mean value of $\mu$ and variance of $\sigma$; $\mathcal{J}(a,b)$ is Jeffreys (log-uniform) prior with lower and upper bounds $a, b$; $\mathcal{U}(a,b)$ is uniform prior with lower and upper bounds $a, b$.
}
\label{tab:rv_gp_result}
\begin{tabular*}{\textwidth}{l@{\extracolsep{\fill}}lc}
\hline
\hline
Parameters & Prior & Posterior \\
\hline
\multicolumn{3}{l}{\textbf{Planet Parameters}} \\
$P_b$ [day]                     & $\mathcal{N}(10.355230,1\times10^{-5})$           & $10.35523 \pm 0.00001$ \\
$T_{\rm b, conj}$ [BJD-2450000] & $\mathcal{N}(8950.8289,5\times10^{-4})$           & $8950.82890 \pm 0.00044$ \\
$K_b$ [m/s]                     & $\mathcal{U}(-50,50)$                             & $2.32^{+0.63}_{-0.63}$ \\
$P_c$ [day]                     & $\mathcal{N}(21.01549,3\times10^{-5})$            & $21.01549 \pm 0.00003$ \\
$T_{\rm c, conj}$ [BJD-2450000] & $\mathcal{N}(8937.8283,7\times10^{-4})$           & $8937.82829 \pm 0.00062$ \\
$K_c$ [m/s]                     & $\mathcal{U}(-50,50)$                             & $1.78^{+0.63}_{-0.62}$ \\
$P_d$ [day]                     & $\mathcal{N}(35.12551,7\times10^{-5})$            & $35.12551 \pm 0.00006$ \\
$T_{\rm d, conj}$ [BJD-2450000] & $\mathcal{N}(8938.296,1\times10^{-3})$            & $8938.296008 \pm 0.0009$ \\
$K_d$ [m/s]                     & $\mathcal{U}(-50,50)$                             & $1.32^{+0.59}_{-0.56}$ \\
$P_e$ [day]                     & $\mathcal{N}(3.0223445,3\times10^{-5})$           & $3.02234 \pm 0.00003$ \\
$T_{\rm e, conj}$ [BJD-2450000] & $\mathcal{N}(8740.21306,8\times10^{-3})$          & $8740.2127 \pm 0.0072$ \\
$K_e$ [m/s]                     & $\mathcal{U}(-50,50)$                             & $2.40^{+0.67}_{-0.65}$ \\
\hline
\multicolumn{3}{l}{\textbf{GP Parameters}} \\
$h_{\rm RV}$ [m/s]              & $\mathcal{J}(0.001,1000)$         & $34.60^{+2.16}_{-1.92}$ \\
$\tau$ [day]                    & $\mathcal{J}(0.001,1000)$         & $26.56^{+1.27}_{-1.18}$ \\
$\Gamma$                        & $\mathcal{J}(0.001,1000)$         & $6.07^{+0.64}_{-0.56}$ \\
$P_{\rm rot}$ [day]             & $\mathcal{N}(7.34,0.2)$           & $7.34^{+0.01}_{-0.01}$ \\
\hline
\multicolumn{3}{l}{\textbf{Instrumental Parameters}} \\
$\gamma_{\rm HIRES}$ [m/s]      & $\mathcal{U}(-50.0,50.0)$         & $ 0.10^{+4.14}_{-4.22}$ \\
$\gamma_{\rm APF}$ [m/s]        & $\mathcal{U}(-50.0,50.0)$         & $-7.34^{+4.35}_{-4.45}$ \\
$\gamma_{\rm HARPS-N}$ [m/s]      & $\mathcal{U}(-50.0,50.0)$         & $ 5.38^{+3.90}_{-4.11}$ \\
$\sigma_{\rm HIRES}$ [m/s]      & $\mathcal{J}(0.001,1000)$         & $ 8.15^{+1.31}_{-1.15}$ \\
$\sigma_{\rm APF}$ [m/s]        & $\mathcal{J}(0.001,1000)$         & $11.74^{+1.78}_{-1.62}$ \\
$\sigma_{\rm HARPS-N}$ [m/s]      & $\mathcal{J}(0.001,1000)$         & $ 4.69^{+0.82}_{-0.76}$ \\
\hline
\multicolumn{3}{l}{\textbf{Derived Masses}} \\
Mass b [$M_\oplus$] & & $7.13^{+1.92}_{-1.94}$ \\
Mass c [$M_\oplus$] & & $6.91^{+2.45}_{-2.42}$ \\
Mass d [$M_\oplus$] & & $6.11^{+2.73}_{-2.57}$ \\
Mass e [$M_\oplus$] & & $4.89^{+1.36}_{-1.33}$ \\
\hline
\end{tabular*}

\end{table}

\begin{figure}[!h]
    \centering
    \includegraphics[width=0.7\linewidth]{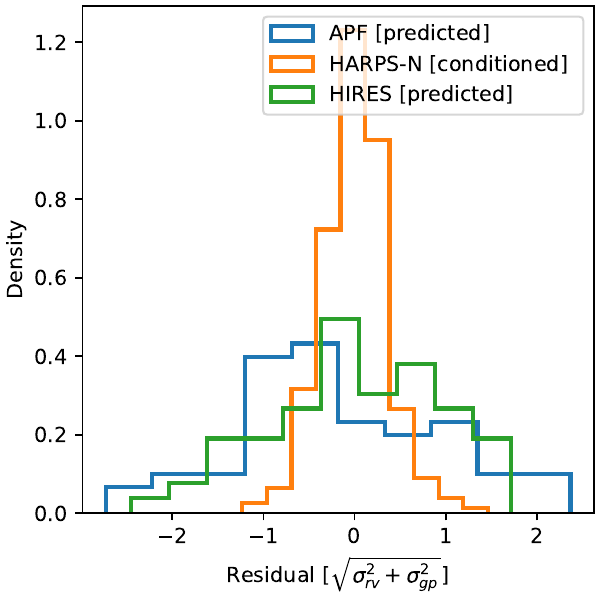}
    \caption{\textbf{Cross-validation test of GP model.} The best-fit GP model is conditioned on HARPS-N data and used to predict the contemporaneous HIRES and APF RV data. The histogram shows the residual of the RV data subtracted by GP model, which are given in unit of the quadracture sum of RV measurement error and GP standard deviation.}
    \label{fig:cross_validation}
\end{figure}

\begin{figure}[!h]
    \centering
    \includegraphics[width=0.95\linewidth]{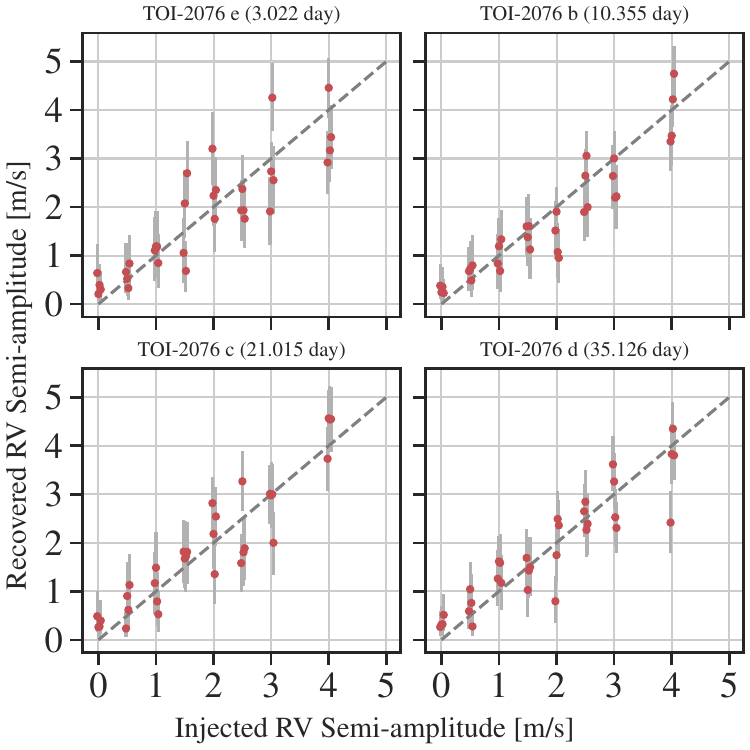}
    \caption{\textbf{The accuracy of retrieved semi-amplitudes of planetary RV signals from GP model.} The median and 68\% credible intervals of the recovered planetary RV semi-amplitudes as a function of the injected semi-amplitudes, based on simulated datasets for the four planets in the system.}
    \label{fig:inj_rec}
\end{figure}

\begin{figure}[!h]
    \centering
    \includegraphics[width=0.95\linewidth]{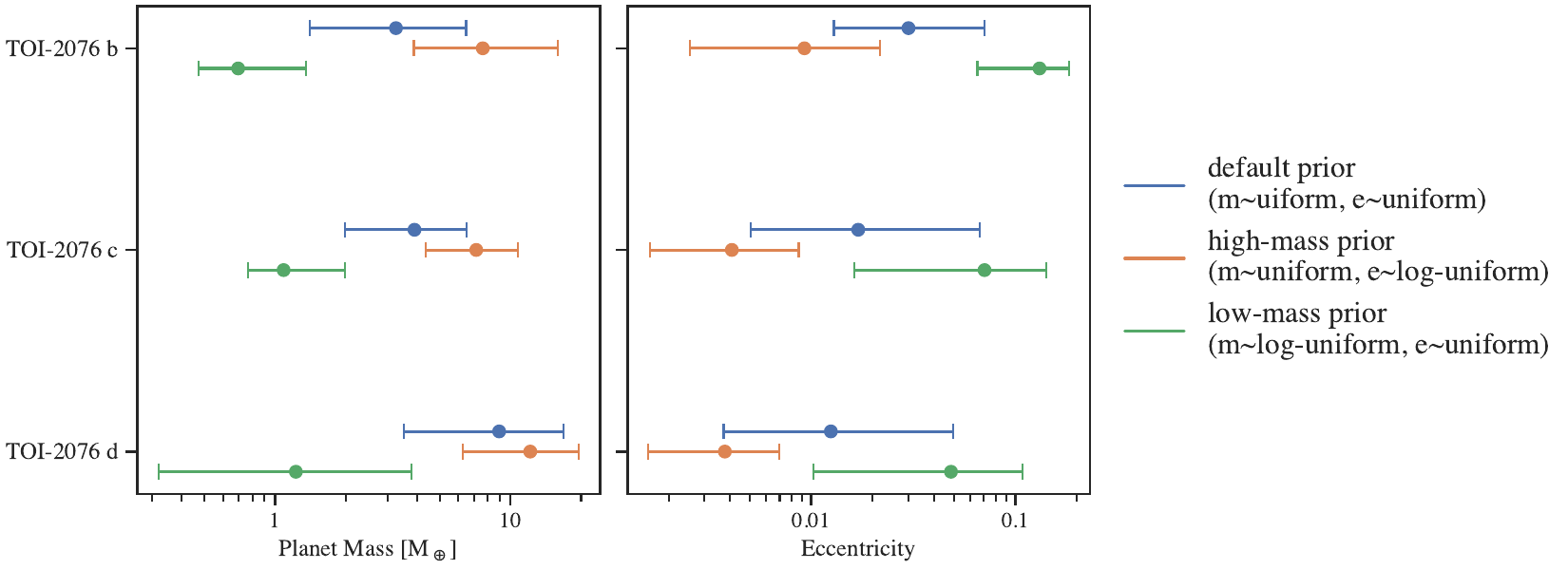}
    \caption{\textbf{Comparisons of planetary mass and eccentricity from TTV analysis with different priors.} The errorbar shows the median and 68\% credible interval of each parameter. These parameters constrained by TTV data alone are highly degenerate and is sensitive to adopted priors.}
    \label{fig:ttv_mass_ecc}
\end{figure}

\begin{figure}[!h]
	\centering
	\includegraphics[width=0.9\textwidth]{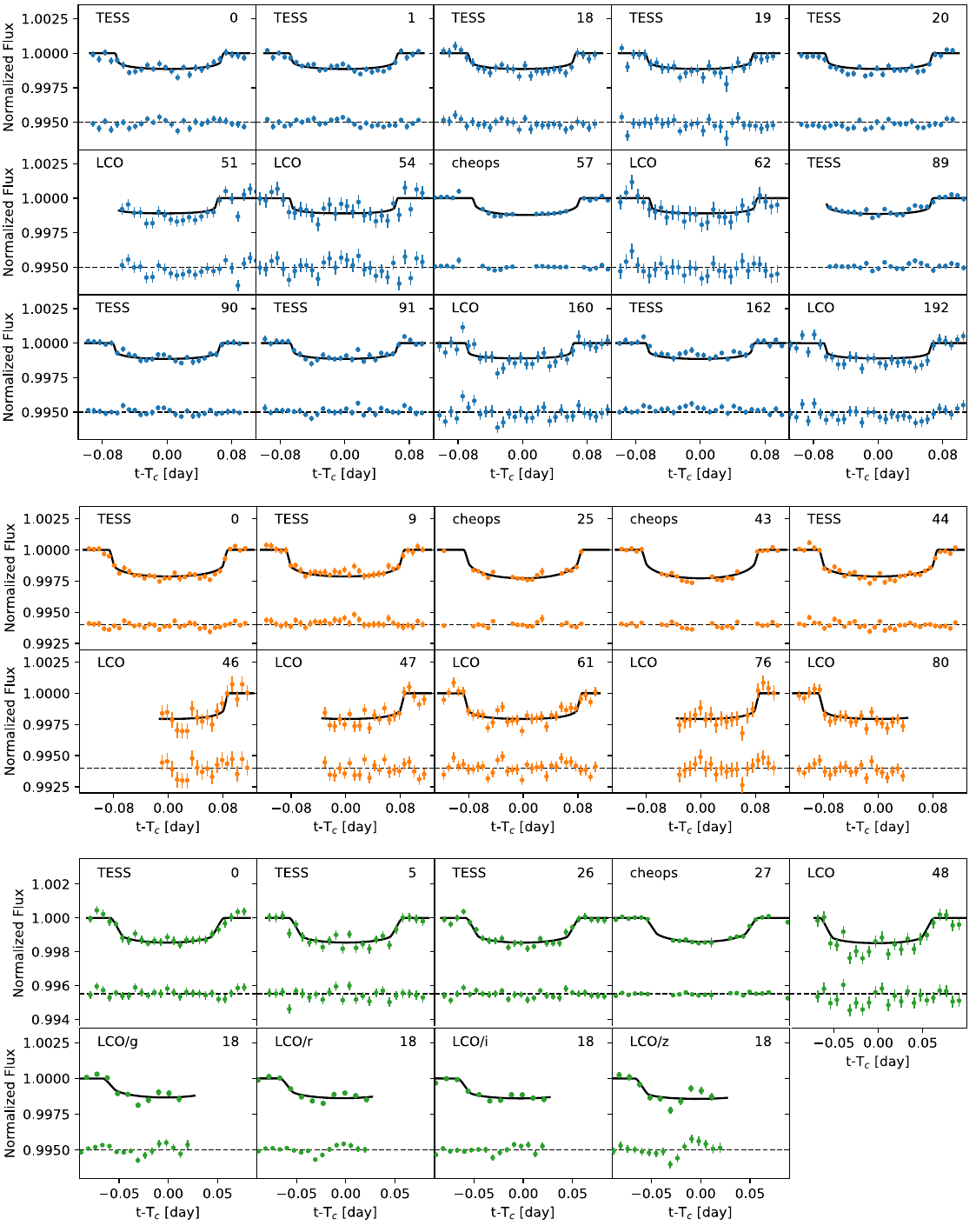} % for an image file named example_figure.*
	% Pick an appriopriate width for the size of the image

	% Captions go below figures
	\caption{\textbf{Best-fit photodynamical model for light curves.}
        The observed transit light curves of TOI-2076 b, c, d are shown in blue, orange and green, respectively. Errorbars are the median and 68\% credible interval of photometric data binned to 10-min
        and shifted to the transit midtimes listed in Table \ref{tab:obs_midtime}. The instruments and transit epoch are shown in the upper part of each panel.
        The black line is the best-fit photodynamical model. Residuals are plotted in the lower part of the figure.
    }
	\label{fig:bestfit_photodynamical_model} % give each figure a logical label name
\end{figure}

\begin{table}
%\fontsize{11pt}{11pt}\selectfont
\centering
\caption{\textbf{Posterior median and 68.3\% (1$\sigma$) credible intervals of fitted planetary orbit parameters from the TTV-only model
and joint photodynamical and RV model.} Osculating Jacobian orbital parameters valid at BJD=2458743. Time of Conjunction defined in t$_{\rm BJD}=2457000$. $\mathcal{N}(\mu,\sigma)$ is normal prior centered at $\mu$ with variance of $\sigma$; $\mathcal{U}(a,b)$ is uniform prior with lower and upper bounds a, b.}
\label{tab:model_bestfit}
\begin{tabular*}{\textwidth}{l@{\extracolsep{\fill}}lll}
\hline\hline
Parameter & Prior & TTV & Photodynamical+RV  \\ 
\hline
Mass Parameters & & \\
Mass ratio, b $M_b/M_s$	($\times10^{-5}$)			& 	$\mathcal{U}(0.1,10)$		& 	 $2.2^{+2.0}_{-1.3}$  	    & 	 $2.39^{+0.66}_{-0.64}$  	\\ 
Mass ratio, c $M_c/M_s$	($\times10^{-5}$)			& 	$\mathcal{U}(0.1,10)$		& 	 $2.2^{+1.1}_{-1.1}$  	    & 	 $2.58^{+0.50}_{-0.52}$  	\\ 
Mass ratio, d $M_d/M_s$	($\times10^{-5}$)			& 	$\mathcal{U}(0.1,10)$		& 	 $4.2^{+2.6}_{-2.3}$        & 	 $2.60^{+0.97}_{-0.98}$  	\\ 
Mass ratio, e $M_e/M_s$	($\times10^{-5}$)			& 	$\mathcal{U}(0.1,10)$		& 	          -                 & 	 $1.69^{+0.53}_{-0.53}$  	\\ 
\hline
Planet b orbit & & \\
Orbital period         $P_b$ (day)				    & 	$\mathcal{N}(10.355,0.1)$ 	& 	 $10.35516^{+0.00032}_{-0.00033}$   & 	 $10.35504^{+0.00015}_{-0.00013}$  	\\  
Eccentricity parameter $e_b\cos\omega_b$	        & 	$\mathcal{U}(-0.1,0.1)	$	& 	 $0.0058^{+0.0105}_{-0.0052}$  	    & 	 $0.0051^{+0.0049}_{-0.0036}$  		\\ 
Eccentricity parameter $e_b\sin\omega_b$	        & 	$\mathcal{U}(-0.1,0.1)	$	& 	 $0.0113^{+0.0216}_{-0.0089}$  	    & 	 $0.009^{+0.0061}_{-0.0044}$  		\\ 
Impact parameter       $b_b$						& 	$\mathcal{U}(0,1)		$	&     -                                 & 	 $0.107^{+0.084}_{-0.072}$  		\\  
Time of conjunction    $T_{\rm conj,b}$ (day)		& 	$\mathcal{N}(1743.7,0.1)$	&    $1743.72067^{+0.00091}_{-0.00094}$	& 	 $1743.72001^{+0.00086}_{-0.00093}$ \\ 
\hline 
Planet c orbit & & \\
Orbital period         $P_c$					    & 	$\mathcal{N}(21.014,0.1)$ 	& 	 $21.0133^{+0.0012}_{-0.0013}$      & 	 $21.01447^{+0.00056}_{-0.00063}$  	\\ 
Eccentricity parameter $e_c\cos\omega_c$	        & 	$\mathcal{U}(-0.1,0.1) 	$	& 	 $0.0008^{+0.014}_{-0.0087}$  	    & 	 $-0.0011^{+0.0089}_{-0.0079}$  	\\ 
Eccentricity parameter $e_c\sin\omega_c$	        & 	$\mathcal{U}(-0.1,0.1)	$	& 	 $0.0004^{+0.0144}_{-0.0094}$  	    & 	 $0.0019^{+0.0095}_{-0.0076}$  		\\ 
Impact parameter       $b_c$						& 	$\mathcal{U}(0,1)		$	&     -                                 & 	 $0.169^{+0.071}_{-0.089}$  		\\ 
Time of conjunction    $T_{\rm conj,c}$			    & 	$\mathcal{N}(1748.7,0.1)$	& 	 $1748.69381^{+0.00071}_{-0.00074}$	& 	 $1748.69359^{+0.00076}_{-0.00074}$ \\ 
\hline
Planet d orbit & & \\
Orbital period         $P_d$					    & 	$\mathcal{N}(35.127,0.1)$ 	& 	 $35.1279^{+0.0012}_{-0.0011}$      & 	 $35.12803^{+0.0006}_{-0.00058}$  	\\ 
Eccentricity parameter $e_d\cos\omega_d$	        & 	$\mathcal{U}(-0.1,0.1) 	$	& 	 $0.0001^{+0.0093}_{-0.0075}$  	    & 	 $0.0011^{+0.0069}_{-0.0058}$  		\\ 
Eccentricity parameter $e_d\sin\omega_d$	        & 	$\mathcal{U}(-0.1,0.1)	$	& 	 $0.0006^{+0.0112}_{-0.0068}$  	    & 	 $-0.0004^{+0.007}_{-0.0065}$  		\\ 
Impact parameter       $b_d$						& 	$\mathcal{U}(0,1)		$	&     -                                 & 	 $0.8157^{+0.0069}_{-0.0064}$  		\\ 
Time of conjunction    $T_{\rm conj,d}$			    & 	$\mathcal{N}(1762.6,0.1)$	& 	 $1762.6669^{+0.0014}_{-0.0014}$	& 	 $1762.6667^{+0.0015}_{-0.0015}$  	\\ 
\hline
Planet e orbit & & \\
Orbital period         $P_e$					    & 	$\mathcal{N}(3.022,0.1)$ 	& 	 -      & 	 $3.0223753^{+8.2e-06}_{-7.7e-06}$  \\ 
Impact parameter       $b_e$						& 	$\mathcal{U}(0,1)		$	&    -      & 	 $0.102^{+0.086}_{-0.071}$  		\\ 
Time of conjunction    $T_{\rm conj,e}$			    & 	$\mathcal{N}(1740.2,0.1)$	& 	 -	    & 	 $1740.21228^{+0.00056}_{-0.00057}$ \\ 
\hline
\end{tabular*}
\end{table}

\begin{table}
\centering
\caption{\textbf{Posterior median and 68.3\% (1$\sigma$) credible intervals of fitted planetary parameters from the joint photodynamical and RV model.}}
\label{tab:model_bestfit1}
\begin{tabular*}{\textwidth}{l@{\extracolsep{\fill}}lc}
\hline\hline
Parameter & Prior & Posterior  \\ 
\hline
Stellar parameters & & \\
Stellar radius         $R_s$ (R$_\odot$)			& 	$\mathcal{N}(0.758,0.014)$    	& 	 $0.7636^{+0.0091}_{-0.0086}$  		\\ 
Stellar mass           $M_s$ (M$_\odot$)			& 	$\mathcal{N}(0.849,0.026)$    	& 	 $0.842^{+0.024}_{-0.024}$  		\\ 
\hline
Planet radii & & \\
Radius ratio, b        $R_b/R_s$					& 	$\mathcal{N}(0.031,0.005)$ 	    & 	 $0.03131^{+0.00025}_{-0.00025}$  	\\ 
Radius ratio, c        $R_c/R_s$					& 	$\mathcal{N}(0.042,0.005)$ 	    & 	 $0.04288^{+0.00031}_{-0.0003}$  	\\ 
Radius ratio, d        $R_d/R_s$					& 	$\mathcal{N}(0.039,0.005)$ 	    & 	 $0.03953^{+0.00038}_{-0.00037}$  	\\ 
Radius ratio, e        $R_e/R_s$					& 	$\mathcal{N}(0.015,0.005)$ 	    & 	 $0.01575^{+0.00035}_{-0.00036}$  	\\ 
\hline
RV parameters && \\
Activity amplitude $h_{RV}$	(m/s)				        & 	$\mathcal{J}(0.001,1000)$	& 	 $34.5^{+2.4}_{-2.1}$  				\\ 
Decay timescale $\tau$	(day)					        & 	$\mathcal{J}(0.001,1000)$	& 	 $26.6^{+1.4}_{-1.3}$  				\\ 
Additive factor  $\Gamma$					            & 	$\mathcal{J}(0.001,1000)$	& 	 $6.1^{+0.72}_{-0.63}$  			\\ 
Rotation period  $T$ 		(day)		                & 	$\mathcal{N}(7.34,0.2)  $	& 	 $7.338^{+0.012}_{-0.012}$  		\\ 
HIRES RV offset $\gamma_{\rm hires}$ (m/s)		        & 	$\mathcal{U}(-50.0,50.0)$	& 	 $0.2^{+4.7}_{-4.6}$  				\\ 
APF RV offset $\gamma_{\rm apf}$ (m/s)			        & 	$\mathcal{U}(-50.0,50.0)$	& 	 $-7.2^{+4.9}_{-4.9}$  				\\ 
HARPS-N RV offset $\gamma_{\rm HARPS-N}$ (m/s)		    & 	$\mathcal{U}(-50.0,50.0)$	& 	 $5.4^{+4.5}_{-4.4}$  				\\ 
HIRES RV jitter noise $\sigma_{\rm hires}$(m/s)		    & 	$\mathcal{J}(0.001,1000)$	& 	 $8.1^{+1.4}_{-1.3}$  				\\ 
APF RV jitter noise $\sigma_{\rm apf}$	(m/s)	        & 	$\mathcal{J}(0.001,1000)$	& 	 $11.9^{+2.1}_{-1.8}$  				\\ 
HARPS-N RV jitter noise $\sigma_{\rm HARPS-N}$(m/s)		& 	$\mathcal{J}(0.001,1000)$	& 	 $4.76^{+0.92}_{-0.91}$  			\\ 
\hline
Limb-darkening Coefficients && \\
$q_{\rm 1,TESS}$			                        & 	$\mathcal{U}(0,1)$			 & 	 $0.298^{+0.105}_{-0.088}$  		\\ 
$q_{\rm 2,TESS}$			                        & 	$\mathcal{U}(0,1)$			 & 	 $0.47^{+0.14}_{-0.13}$  			\\ 
$q_{\rm 1,z}$				                        & 	$\mathcal{U}(0,1)$			 & 	 $0.114^{+0.141}_{-0.081}$  		\\ 
$q_{\rm 2,z}$				                        & 	$\mathcal{U}(0,1)$			 & 	 $0.359^{+0.107}_{-0.087}$  		\\ 
$q_{\rm 1,CHEOPS}$			                        & 	$\mathcal{U}(0,1)$			 & 	 $0.67^{+0.15}_{-0.13}$  			\\  
$q_{\rm 2,CHEOPS}$			                        & 	$\mathcal{U}(0,1)$			 & 	 $0.31^{+0.18}_{-0.14}$  			\\  
\hline
\end{tabular*}
\end{table}

\begin{figure} % Do not use \begin{figure*}
	\centering
	\includegraphics[width=1.0\textwidth]{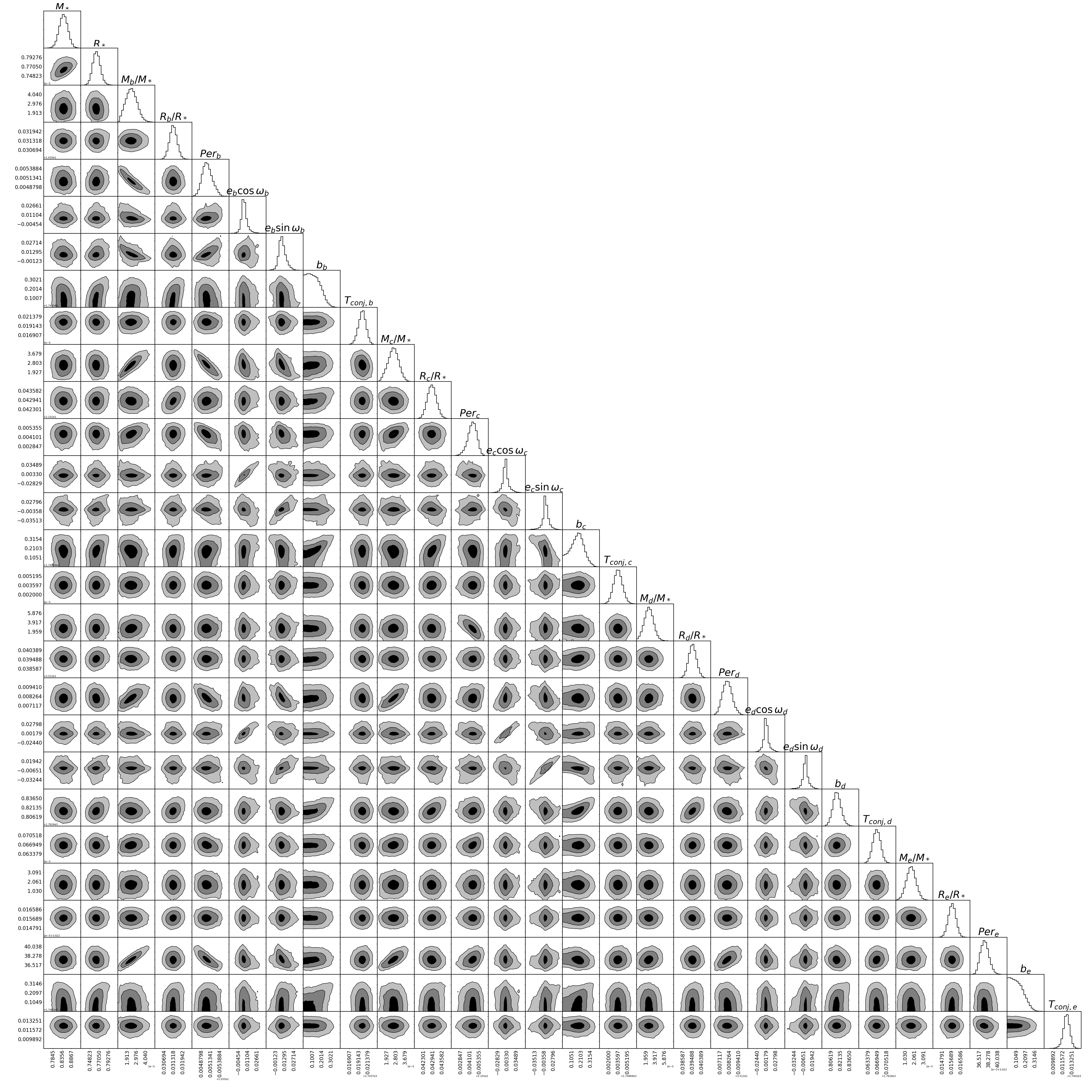} % for an image file named example_figure.*
	% Pick an appriopriate width for the size of the image
	% Captions go below figures
	\caption{\textbf{Cornerplots of fitted stellar andplanetary parameters in joint photodynamical+RV model.} The black contours on the
    2D panels represent the 1$\sigma$, 2$\sigma$, and 3$\sigma$ confidence levels of the overall posterior samples. The diagonal panels show the marginalized probability distribution of each parameter}
	\label{fig:corner_plot} % give each figure a logical label name
\end{figure}

\backmatter





\newpage 

\begin{thebibliography}{104}
\ifx \bisbn   \undefined \def \bisbn  #1{ISBN #1}\fi
\ifx \binits  \undefined \def \binits#1{#1}\fi
\ifx \bauthor  \undefined \def \bauthor#1{#1}\fi
\ifx \batitle  \undefined \def \batitle#1{#1}\fi
\ifx \bjtitle  \undefined \def \bjtitle#1{\textit{#1}}\fi
\ifx \bvolume  \undefined \def \bvolume#1{#1}\fi
\ifx \byear  \undefined \def \byear#1{#1}\fi
\ifx \bissue  \undefined \def \bissue#1{#1}\fi
\ifx \bfpage  \undefined \def \bfpage#1{#1}\fi
\ifx \blpage  \undefined \def \blpage #1{#1}\fi
\ifx \burl  \undefined \def \burl#1{\textsf{#1}}\fi
\ifx \doiurl  \undefined \def \doiurl#1{\url{https://doi.org/#1}}\fi
\ifx \betal  \undefined \def \betal{~et al.~}\fi
\ifx \binstitute  \undefined \def \binstitute#1{#1}\fi
\ifx \binstitutionaled  \undefined \def \binstitutionaled#1{#1}\fi
\ifx \bctitle  \undefined \def \bctitle#1{#1}\fi
\ifx \beditor  \undefined \def \beditor#1{#1}\fi
\ifx \bpublisher  \undefined \def \bpublisher#1{#1}\fi
\ifx \bbtitle  \undefined \def \bbtitle#1{\textit{#1}}\fi
\ifx \bedition  \undefined \def \bedition#1{#1}\fi
\ifx \bseriesno  \undefined \def \bseriesno#1{#1}\fi
\ifx \blocation  \undefined \def \blocation#1{#1}\fi
\ifx \bsertitle  \undefined \def \bsertitle#1{#1}\fi
\ifx \bsnm \undefined \def \bsnm#1{#1}\fi
\ifx \bsuffix \undefined \def \bsuffix#1{#1}\fi
\ifx \bparticle \undefined \def \bparticle#1{#1}\fi
\ifx \barticle \undefined \def \barticle#1{#1}\fi
\bibcommenthead
\ifx \bconfdate \undefined \def \bconfdate #1{#1}\fi
\ifx \botherref \undefined \def \botherref #1{#1}\fi
\ifx \url \undefined \def \url#1{\textsf{#1}}\fi
\ifx \bchapter \undefined \def \bchapter#1{#1}\fi
\ifx \bbook \undefined \def \bbook#1{#1}\fi
\ifx \bcomment \undefined \def \bcomment#1{#1}\fi
\ifx \oauthor \undefined \def \oauthor#1{#1}\fi
\ifx \citeauthoryear \undefined \def \citeauthoryear#1{#1}\fi
\ifx \endbibitem  \undefined \def \endbibitem {}\fi
\ifx \bconflocation  \undefined \def \bconflocation#1{#1}\fi
\ifx \arxivurl  \undefined \def \arxivurl#1{\textsf{#1}}\fi
\csname PreBibitemsHook\endcsname

%%% 1
\bibitem[\protect\citeauthoryear{{Terquem} and {Papaloizou}}{2007}]{TerquemPapaloizou2007}
\begin{barticle}
\bauthor{\bsnm{{Terquem}}, \binits{C.}},
\bauthor{\bsnm{{Papaloizou}}, \binits{J.C.B.}}
\batitle{{Migration and the Formation of Systems of Hot Super-Earths and Neptunes}}.
\bjtitle{Astrophys. J.}
\bvolume{654}(\bissue{2}),
\bfpage{1110}--\blpage{1120}
(\byear{2007})
\end{barticle}
\endbibitem

%%% 2
\bibitem[\protect\citeauthoryear{{Murray} and {Dermott}}{1999}]{MurrayDermott}
\begin{bbook}
\bauthor{\bsnm{{Murray}}, \binits{C.D.}},
\bauthor{\bsnm{{Dermott}}, \binits{S.F.}}
\bbtitle{{Solar System Dynamics}}
(\byear{Cambridge University Press, 1999})
\end{bbook}
\endbibitem

%%% 3
\bibitem[\protect\citeauthoryear{Izidoro et~al.}{2017}]{Izidoro2017}
\begin{barticle}
\bauthor{\bsnm{Izidoro}, \binits{A.}}\betal
\batitle{{Breaking the chains: Hot super-Earth systems from migration and disruption of compact resonant chains}}.
\bjtitle{Mon. Not. R. Astron. Soc.}
\bvolume{470}(\bissue{2}),
\bfpage{1750}--\blpage{1770}
(\byear{2017})
\end{barticle}
\endbibitem

%%% 4
\bibitem[\protect\citeauthoryear{{Fabrycky} et~al.}{2014}]{Fabrycky2014}
\begin{barticle}
\bauthor{\bsnm{{Fabrycky}}, \binits{D.C.}}\betal
\batitle{{Architecture of Kepler's Multi-transiting Systems. II. New Investigations with Twice as Many Candidates}}.
\bjtitle{Astrophys. J.}
\bvolume{790}(\bissue{2}),
\bfpage{146}
(\byear{2014})
\end{barticle}
\endbibitem

%%% 5
\bibitem[\protect\citeauthoryear{{Dai} et~al.}{2024}]{Dai2024}
\begin{barticle}
\bauthor{\bsnm{{Dai}}, \binits{F.}} \betal
\batitle{{The Prevalence of Resonance Among Young, Close-in Planets}}.
\bjtitle{Astron. J.}
\bvolume{168}(\bissue{6}),
\bfpage{239}
(\byear{2024})
\end{barticle}
\endbibitem


%%% 6
\bibitem[\protect\citeauthoryear{Hamer and Schlaufman}{2024}]{Hamer2024}
\begin{barticle}
\bauthor{\bsnm{Hamer}, \binits{J.H.}},
\bauthor{\bsnm{Schlaufman}, \binits{K.C.}}
\batitle{{Kepler-discovered Multiple-planet Systems near Period Ratios Suggestive of Mean-motion Resonances Are Young}}.
\bjtitle{Astron. J.}
\bvolume{167}(\bissue{2}),
\bfpage{55}
(\byear{2024})
\end{barticle}
\endbibitem

%%% 7
\bibitem[\protect\citeauthoryear{Huang and Ormel}{2023}]{huang2023and}
\begin{barticle}
\bauthor{\bsnm{Huang}, \binits{S.}},
\bauthor{\bsnm{Ormel}, \binits{C.W.}}
\batitle{When, where, and how many planets end up in first-order resonances?}
\bjtitle{Mon. Not. R. Astron. Soc.}
\bvolume{522}(\bissue{1}),
\bfpage{828}--\blpage{846}
(\byear{2023})
\end{barticle}
\endbibitem

%%% 8
\bibitem[\protect\citeauthoryear{{Goldberg}, {Batygin} and {Morbidelli}}{2022}]{Goldberg_stability}
\begin{barticle}
\bauthor{\bsnm{{Goldberg}}, \binits{M.}},
\bauthor{\bsnm{{Batygin}}, \binits{K.}},
\bauthor{\bsnm{{Morbidelli}}, \binits{A.}}
\batitle{{A criterion for the stability of planets in chains of resonances}}.
\bjtitle{Icarus}
\bvolume{388},
\bfpage{115206}
(\byear{2022})
\end{barticle}
\endbibitem


%%% 9
\bibitem[\protect\citeauthoryear{{Nesvorn{\'y}}}{2018}]{Nesvorny2018}
\begin{barticle}
\bauthor{\bsnm{{Nesvorn{\'y}}}, \binits{D.}}
\batitle{{Dynamical Evolution of the Early Solar System}}.
\bjtitle{Annu. Rev. Astron. Astrophys.}
\bvolume{56},
\bfpage{137}--\blpage{174}
(\byear{2018})
\end{barticle}
\endbibitem

%%% 10
\bibitem[\protect\citeauthoryear{{Hedges} et~al.}{2021}]{Hedges2021}
\begin{barticle}
\bauthor{\bsnm{{Hedges}}, \binits{C.}}\betal
\batitle{{TOI-2076 and TOI-1807: Two Young, Comoving Planetary Systems within 50 pc Identified by TESS that are Ideal Candidates for Further Follow Up}}.
\bjtitle{Astron. J.}
\bvolume{162}(\bissue{2}),
\bfpage{54}
(\byear{2021})
\end{barticle}
\endbibitem

%%% 11
\bibitem[\protect\citeauthoryear{{Barber}, {Mann}, {Vanderburg}, {Boyle} and {Lopez Murillo}}{2025}]{Barber2025}
\begin{barticle}
\bauthor{\bsnm{{Barber}}, \binits{M.G.}},
\bauthor{\bsnm{{Mann}}, \binits{A.W.}},
\bauthor{\bsnm{{Vanderburg}}, \binits{A.}},
\bauthor{\bsnm{{Boyle}}, \binits{A.W.}},
\bauthor{\bsnm{{Lopez Murillo}}, \binits{A.I.}}
\batitle{{TESS Investigation—Demographics of Young Exoplanets (TI-DYE). III. An Inner Super-Earth in TOI~2076}}.
\bjtitle{Astron. J.}
\bvolume{170}(\bissue{1}),
\bfpage{32}
(\byear{2025})
\end{barticle}
\endbibitem

%%% 12
\bibitem[\protect\citeauthoryear{{Frazier} et~al.}{2023}]{Frazier}
\begin{barticle}
\bauthor{\bsnm{{Frazier}}, \binits{R.C.}}\betal
\batitle{{NEID Reveals That the Young Warm Neptune TOI-2076 b Has a Low Obliquity}}.
\bjtitle{Astrophys. J.}
\bvolume{944}(\bissue{2}),
\bfpage{41}
(\byear{2023})
\end{barticle}
\endbibitem

%%% 13
\bibitem[\protect\citeauthoryear{Lithwick et~al.}{2012}]{Lithwick2012_ttv}
\begin{barticle}
\bauthor{\bsnm{Lithwick}, \binits{Y.}},
\bauthor{\bsnm{Xie}, \binits{J.}},
\bauthor{\bsnm{Wu}, \binits{Y.}}
\batitle{Extracting planet mass and eccentricity from TTV data}.
\bjtitle{Astrophys. J.}
\bvolume{761}(\bissue{2}),
\bfpage{122}
(\byear{2012})
\end{barticle}
\endbibitem
%%% 14
\bibitem[\protect\citeauthoryear{Nesvorn{\'{y}} and Vokrouhlick{\'{y}}}{2016}]{Nesvorny2016}
\begin{barticle}
\bauthor{\bsnm{Nesvorn{\'{y}}}, \binits{D.}},
\bauthor{\bsnm{Vokrouhlick{\'{y}}}, \binits{D.}}
\batitle{{Dynamics and Transit Variations of Resonant Exoplanets}}.
\bjtitle{Astrophys. J.}
\bvolume{823}(\bissue{2}),
\bfpage{72}
(\byear{2016})
\end{barticle}
\endbibitem

%%% 15
\bibitem[\protect\citeauthoryear{{Rafikov}}{2006}]{Rafikov}
\begin{barticle}
\bauthor{\bsnm{{Rafikov}}, \binits{R.R.}}
\batitle{{Atmospheres of Protoplanetary Cores: Critical Mass for Nucleated Instability}}.
\bjtitle{Astrophys. J.}
\bvolume{648},
\bfpage{666}--\blpage{682}
(\byear{2006})
\end{barticle}
\endbibitem

%%% 16
\bibitem[\protect\citeauthoryear{Chen and Rogers}{2016}]{Chen2016}
\begin{barticle}
\bauthor{\bsnm{Chen}, \binits{H.}},
\bauthor{\bsnm{Rogers}, \binits{L.A.}}
\batitle{{Evolutionary Analysis of Gaseous Sub-Neptune-Mass Planets With Mesa}}.
\bjtitle{Astrophys. J.}
\bvolume{831}(\bissue{2}),
\bfpage{180}
(\byear{2016})
\end{barticle}
\endbibitem

%%% 17
\bibitem[\protect\citeauthoryear{Zeng et~al.}{2019}]{Zeng2019}
\begin{barticle}
\bauthor{\bsnm{Zeng}, \binits{L.}}\betal
\batitle{Growth model interpretation of planet size distribution}.
\bjtitle{Proc. Natl Acad. Sci. USA}
\bvolume{116}(\bissue{20}),
\bfpage{9723}--\blpage{9728}
(\byear{2019})
\end{barticle}
\endbibitem

%%% 18
\bibitem[\protect\citeauthoryear{{Aguichine} et~al.}{2025}]{Aguichine2024}
\begin{barticle}
\bauthor{\bsnm{{Aguichine}}, \binits{A.}} \betal
\batitle{{Evolution of Steam Worlds: Energetic Aspects}}.
\bjtitle{Astrophys. J.}
\bvolume{988}(\bissue{2}),
\bfpage{186}
(\byear{2025})
\end{barticle}
\endbibitem

%%% 19
\bibitem[\protect\citeauthoryear{{Leleu} et~al.}{2024}]{Leleu2024}
\begin{barticle}
\bauthor{\bsnm{{Leleu}}, \binits{A.}}\betal
\batitle{{Resonant sub-Neptunes are puffier}}.
\bjtitle{Astron. Astrophys.}
\bvolume{687},
\bfpage{1}
(\byear{2024})
\end{barticle}
\endbibitem

%%% 20
\bibitem[\protect\citeauthoryear{Dai et~al.}{2023}]{Dai2023}
\begin{barticle}
\bauthor{\bsnm{Dai}, \binits{F.}}\betal
\batitle{{TOI-1136 is a Young, Coplanar, Aligned Planetary System in a Pristine Resonant Chain}}.
\bjtitle{Astron. J.}
\bvolume{165}(\bissue{2}),
\bfpage{33}
(\byear{2023})
\end{barticle}
\endbibitem

%%% 21
\bibitem[\protect\citeauthoryear{Hadden}{2019}]{Hadden2019}
\begin{barticle}
\bauthor{\bsnm{Hadden}, \binits{S.}}
\batitle{{An Integrable Model for the Dynamics of Planetary Mean-motion Resonances}}.
\bjtitle{Astron. J.}
\bvolume{158}(\bissue{6}),
\bfpage{238}
(\byear{2019})
\end{barticle}
\endbibitem

%%% 22
\bibitem[\protect\citeauthoryear{{Keller}, {Dai} and {Xu}}{2026}]{Keller2025}
\begin{barticle}
\bauthor{\bsnm{{Keller}}, \binits{F.M.}},
\bauthor{\bsnm{{Dai}}, \binits{F.}},
\bauthor{\bsnm{{Xu}}, \binits{W.}}:
\batitle{{Higher-order Mean-motion Resonances Can Form in Type I Disk Migration}}.
\bjtitle{Astrophys. J.}
\bvolume{996}(\bissue{1}),
\bfpage{12}
(\byear{2026})
\end{barticle}
\endbibitem


%%% 23
\bibitem[\protect\citeauthoryear{{Batygin}}{2015}]{Batygin2015_capture}
\begin{barticle}
\bauthor{\bsnm{{Batygin}}, \binits{K.}}
\batitle{{Capture of planets into mean-motion resonances and the origins of extrasolar orbital architectures}}.
\bjtitle{Mon. Not. R. Astron. Soc.}
\bvolume{451}(\bissue{3}),
\bfpage{2589}--\blpage{2609}
(\byear{2015})
\end{barticle}
\endbibitem

%%% 24
\bibitem[\protect\citeauthoryear{{Liu} et~al.}{2017}]{Liu2017}
\begin{barticle}
\bauthor{\bsnm{{Liu}}, \binits{B.}},
\bauthor{\bsnm{{Ormel}}, \binits{C.W.}},
\bauthor{\bsnm{{Lin}}, \binits{D.N.C.}}
\batitle{{Dynamical rearrangement of super-Earths during disk dispersal. I. Outline of the magnetospheric rebound model}}.
\bjtitle{Astron. Astrophys.}
\bvolume{601},
\bfpage{15}
(\byear{2017})
\end{barticle}
\endbibitem

%%% 25
\bibitem[\protect\citeauthoryear{{Wu}, {Malhotra} and {Lithwick}}{2024}]{Wu2024}
\begin{barticle}
\bauthor{\bsnm{{Wu}}, \binits{Y.}},
\bauthor{\bsnm{{Malhotra}}, \binits{R.}},
\bauthor{\bsnm{{Lithwick}}, \binits{Y.}}:
\batitle{{Repelling Planet Pairs by Ping-pong Scattering}}.
\bjtitle{Astrophys. J.}
\bvolume{971}(\bissue{1}),
\bfpage{5}
(\byear{2024})
\end{barticle}
\endbibitem


%%% 26
\bibitem[\protect\citeauthoryear{{Wang} and {Lin}}{2023}]{Wang_lin_2023}
\begin{barticle}
\bauthor{\bsnm{{Wang}}, \binits{S.}},
\bauthor{\bsnm{{Lin}}, \binits{D.N.C.}}
\batitle{{Dynamical Evolution of Closely Packed Multiple Planetary Systems Subject to Atmospheric Mass Loss}}.
\bjtitle{Astron. J.}
\bvolume{165}(\bissue{4}),
\bfpage{174}
(\byear{2023})
\end{barticle}
\endbibitem

%%% 27
\bibitem[\protect\citeauthoryear{{Williams} and {Benson}}{1971}]{Williams}
\begin{barticle}
\bauthor{\bsnm{{Williams}}, \binits{J.G.}},
\bauthor{\bsnm{{Benson}}, \binits{G.S.}}
\batitle{{Resonances in the Neptune-Pluto System}}.
\bjtitle{Astron. J.}
\bvolume{76},
\bfpage{167}
(\byear{1971})
\end{barticle}
\endbibitem

%%% 28
\bibitem[\protect\citeauthoryear{Obertas et~al.}{2017}]{Obertas2017}
\begin{barticle}
\bauthor{\bsnm{Obertas}, \binits{A.}},
\bauthor{\bsnm{{Van Laerhoven}}, \binits{C.}},
\bauthor{\bsnm{Tamayo}, \binits{D.}}
\batitle{{The stability of tightly-packed, evenly-spaced systems of Earth-mass planets orbiting a Sun-like star}}.
\bjtitle{Icarus}
\bvolume{293},
\bfpage{52}--\blpage{58}
(\byear{2017})
\end{barticle}
\endbibitem

%%% 29
\bibitem[\protect\citeauthoryear{Rath et~al.}{2022}]{Rath2022}
\begin{barticle}
\bauthor{\bsnm{Rath}, \binits{J.}},
\bauthor{\bsnm{Hadden}, \binits{S.}},
\bauthor{\bsnm{Lithwick}, \binits{Y.}}
\batitle{{The Criterion for Chaos in Three-planet Systems}}.
\bjtitle{Astrophys. J.}
\bvolume{932}(\bissue{1}),
\bfpage{61}
(\byear{2022})
\end{barticle}
\endbibitem

%%% 30
\bibitem[\protect\citeauthoryear{{Lammers} and {Winn}}{2024}]{Lammers2024}
\begin{barticle}
\bauthor{\bsnm{{Lammers}}, \binits{C.}},
\bauthor{\bsnm{{Winn}}, \binits{J.N.}}:
\batitle{{The Six-planet Resonant Chain of HD 110067}}.
\bjtitle{Astrophys. J. Lett.}
\bvolume{968}(\bissue{1}),
\bfpage{L12}
(\byear{2024})
\end{barticle}
\endbibitem


%%% 31
\bibitem[\protect\citeauthoryear{{Wisdom}}{1980}]{Wisdom1980}
\begin{barticle}
\bauthor{\bsnm{{Wisdom}}, \binits{J.}}
\batitle{{The resonance overlap criterion and the onset of stochastic behavior in the restricted three-body problem}}.
\bjtitle{Astron. J.}
\bvolume{85},
\bfpage{1122}--\blpage{1133}
(\byear{1980})
\end{barticle}
\endbibitem

%%% 32
\bibitem[\protect\citeauthoryear{{Hu} et~al.}{2025}]{Hu2025}
\begin{barticle}
\bauthor{\bsnm{{Hu}}, \binits{Z.}} \textit{et~al.}:
\batitle{{Unexpected Near-Resonant and Metastable States of Young Multiplanet Systems}}.
\bjtitle{Astrophys. J.}
\bvolume{995}(\bissue{2}),
\bfpage{206}
(\byear{2025})
\end{barticle}
\endbibitem

%%% 33
\bibitem[\protect\citeauthoryear{{Lithwick} and {Wu}}{2011}]{Lithwick2011}
\begin{barticle}
\bauthor{\bsnm{{Lithwick}}, \binits{Y.}},
\bauthor{\bsnm{{Wu}}, \binits{Y.}}
\batitle{{Theory of Secular Chaos and Mercury's Orbit}}.
\bjtitle{Astrophys. J.}
\bvolume{739}(\bissue{1}),
\bfpage{31}
(\byear{2011})
\end{barticle}
\endbibitem

%%% 34
\bibitem[\protect\citeauthoryear{Lee and Chiang}{2015}]{Lee2015}
\begin{barticle}
\bauthor{\bsnm{Lee}, \binits{E.J.}},
\bauthor{\bsnm{Chiang}, \binits{E.}}
\batitle{{To Cool is to Accrete: Analytic Scalings for Nebular Accretion of Planetary Atmospheres}}.
\bjtitle{Astrophys. J.}
\bvolume{811}(\bissue{1}),
\bfpage{41}
(\byear{2015})
\end{barticle}
\endbibitem

%%% 35
\bibitem[\protect\citeauthoryear{{Fulton} et~al.}{2017}]{Fulton}
\begin{barticle}
\bauthor{\bsnm{{Fulton}}, \binits{B.J.}}\betal
\batitle{{The California-Kepler Survey. III. A Gap in the Radius Distribution of Small Planets}}.
\bjtitle{Astron. J.}
\bvolume{154},
\bfpage{109}
(\byear{2017})
\end{barticle}
\endbibitem

%%% 36
\bibitem[\protect\citeauthoryear{Owen and Wu}{2017}]{Owen2017}
\begin{barticle}
\bauthor{\bsnm{Owen}, \binits{J.E.}},
\bauthor{\bsnm{Wu}, \binits{Y.}}
\batitle{{The Evaporation Valley in the Kepler Planets}}.
\bjtitle{Astrophys. J.}
\bvolume{847}(\bissue{1}),
\bfpage{29}
(\byear{2017})
\end{barticle}
\endbibitem

%%% 37
\bibitem[\protect\citeauthoryear{{Luque} and {Pall{\'e}}}{2022}]{Luque_water}
\begin{barticle}
\bauthor{\bsnm{{Luque}}, \binits{R.}},
\bauthor{\bsnm{{Pall{\'e}}}, \binits{E.}}
\batitle{{Density, not radius, separates rocky and water-rich small planets orbiting M dwarf stars}}.
\bjtitle{Science}
\bvolume{377}(\bissue{6611}),
\bfpage{1211}--\blpage{1214}
(\byear{2022})
\end{barticle}
\endbibitem

%%% 38
\bibitem[\protect\citeauthoryear{{Ginzburg} et~al.}{2018}]{Ginzburg}
\begin{barticle}
\bauthor{\bsnm{{Ginzburg}}, \binits{S.}},
\bauthor{\bsnm{{Schlichting}}, \binits{H.E.}},
\bauthor{\bsnm{{Sari}}, \binits{R.}}
\batitle{{Core-powered mass-loss and the radius distribution of small exoplanets}}.
\bjtitle{Mon. Not. R. Astron. Soc.}
\bvolume{476}(\bissue{1}),
\bfpage{759}--\blpage{765}
(\byear{2018})
\end{barticle}
\endbibitem

%%% 39
\bibitem[\protect\citeauthoryear{{Zhang} et~al.}{2023}]{Zhang2023}
\begin{barticle}
\bauthor{\bsnm{{Zhang}}, \binits{M.}}\betal
\batitle{{Detection of Atmospheric Escape from Four Young Mini-Neptunes}}.
\bjtitle{Astron. J.}
\bvolume{165}(\bissue{2}),
\bfpage{62}
(\byear{2023})
\end{barticle}
\endbibitem

%%% 41
\bibitem[\protect\citeauthoryear{{Biersteker} and {Schlichting}}{2019}]{Biersteker}
\begin{barticle}
\bauthor{\bsnm{{Biersteker}}, \binits{J.B.}},
\bauthor{\bsnm{{Schlichting}}, \binits{H.E.}}
\batitle{{Atmospheric mass-loss due to giant impacts: the importance of the thermal component for hydrogen-helium envelopes}}.
\bjtitle{Mon. Not. R. Astron. Soc.}
\bvolume{485}(\bissue{3}),
\bfpage{4454}--\blpage{4463}
(\byear{2019})
\end{barticle}
\endbibitem

%%% 42
\bibitem[\protect\citeauthoryear{{Ribas} et~al.}{2005}]{Ribas2005}
\begin{barticle}
\bauthor{\bsnm{{Ribas}}, \binits{I.}},
\bauthor{\bsnm{{Guinan}}, \binits{E.F.}},
\bauthor{\bsnm{{G{\"u}del}}, \binits{M.}},
\bauthor{\bsnm{{Audard}}, \binits{M.}}
\batitle{{Evolution of the Solar Activity over Time and Effects on Planetary Atmospheres. I. High-Energy Irradiances (1--1700 \AA)}}.
\bjtitle{Astrophys. J.}
\bvolume{622}(\bissue{1}),
\bfpage{680}--\blpage{694}
(\byear{2005})
\end{barticle}
\endbibitem

%%% 43
\bibitem[\protect\citeauthoryear{{Owen} and {Wu}}{2016}]{Owen_boiloff}
\begin{barticle}
\bauthor{\bsnm{{Owen}}, \binits{J.E.}},
\bauthor{\bsnm{{Wu}}, \binits{Y.}}
\batitle{{Atmospheres of Low-mass Planets: The ``Boil-off''}}.
\bjtitle{Astrophys. J.}
\bvolume{817}(\bissue{2}),
\bfpage{107}
(\byear{2016})
\end{barticle}
\endbibitem

%%% 44
\bibitem[\protect\citeauthoryear{Lithwick and Wu}{2012}]{Lithwick_repulsion}
\begin{barticle}
\bauthor{\bsnm{Lithwick}, \binits{Y.}},
\bauthor{\bsnm{Wu}, \binits{Y.}}
\batitle{Resonant repulsion of \textit{Kepler} planet pairs}.
\bjtitle{Astrophys. J. Lett.}
\bvolume{756},
\bfpage{L11}
(\byear{2012})
\end{barticle}
\endbibitem

%%% 45
\bibitem[\protect\citeauthoryear{{Hanf} et~al.}{2025}]{Hanf}
\begin{barticle}
\bauthor{\bsnm{{Hanf}}, \binits{B.}},
\bauthor{\bsnm{{Kincaid}}, \binits{W.}},
\bauthor{\bsnm{{Schlichting}}, \binits{H.}},
\bauthor{\bsnm{{Cappiello}}, \binits{L.}},
\bauthor{\bsnm{{Tamayo}}, \binits{D.}}
\batitle{{Orbital Migration Through Atmospheric Mass Loss}}.
\bjtitle{Astron. J.}
\bvolume{169}(\bissue{1}),
\bfpage{19}
(\byear{2025})
\end{barticle}
\endbibitem

%%% 46
\bibitem[\protect\citeauthoryear{{David} et~al.}{2019}]{David2019}
\begin{barticle}
\bauthor{\bsnm{{David}}, \binits{T.J.}}\betal
\batitle{{Four Newborn Planets Transiting the Young Solar Analog V1298 Tau}}.
\bjtitle{Astrophys. J. Lett.}
\bvolume{885}(\bissue{1}),
\bfpage{L12}
(\byear{2019})
\end{barticle}
\endbibitem

\bibitem[\protect\citeauthoryear{{Livingston} et~al.}{2026}]{Livingston2026}
\begin{barticle}
\bauthor{\bsnm{{Livingston}}, \binits{J.H.}} \betal:
\batitle{{A young progenitor for the most common planetary systems in the Galaxy}}.
\bjtitle{Nature}
\bvolume{649},
\bfpage{310}--\blpage{314}
(\byear{2026})
\end{barticle}
\endbibitem


%%% 47
\bibitem[\protect\citeauthoryear{Leleu et~al.}{2021}]{Leleu2021}
\begin{barticle}
\bauthor{\bsnm{Leleu}, \binits{A.}}\betal
\batitle{Six transiting planets and a chain of Laplace resonances in TOI-178}.
\bjtitle{Astron. Astrophys.}
\bvolume{651},
\bfpage{A26}
(\byear{2021})
\end{barticle}
\endbibitem

%%% 48
\bibitem[\protect\citeauthoryear{{Vach} et~al.}{2024}]{Vach2024}
\begin{barticle}
\bauthor{\bsnm{{Vach}}, \binits{S.}}\betal
\batitle{{The Occurrence of Small, Short-period Planets Younger than 200 Myr with TESS}}.
\bjtitle{Astron. J.}
\bvolume{167}(\bissue{5}),
\bfpage{210}
(\byear{2024})
\end{barticle}
\endbibitem

%%% 49
\bibitem[\protect\citeauthoryear{{Fernandes} et~al.}{2025}]{Fernandes2025}
\begin{barticle}
\bauthor{\bsnm{{Fernandes}}, \binits{R.B.}}\betal
\batitle{{Signatures of Atmospheric Mass Loss and Planet Migration in the Time Evolution of Short-period Transiting Exoplanets}}.
\bjtitle{Astron. J.}
\bvolume{169}(\bissue{4}),
\bfpage{208}
(\byear{2025})
\end{barticle}
\endbibitem

\bibitem[\protect\citeauthoryear{{Dai} et~al.}{2025}]{Dai2025UPiCII}
\begin{botherref}
\oauthor{\bsnm{{Dai}}, \binits{Y.-Z.}} \betal:
{Understanding the Planetary Formation and Evolution in Star Clusters (UPiC)-II: Catalog of planets/candidates in Open Clusters and Moving Groups}.
Preprint at arXiv:2512.07029
(\byear{2025})
\end{botherref}
\endbibitem

%%% 50
\bibitem[\protect\citeauthoryear{{Jenkins} et~al.}{2016}]{Jenkins2016}
\begin{bchapter}
\bauthor{\bsnm{{Jenkins}}, \binits{J.M.}}\betal
\bctitle{{The TESS science processing operations center}}.
In: \beditor{\bsnm{{Chiozzi}}, \binits{G.}},
\beditor{\bsnm{{Guzman}}, \binits{J.C.}} (eds.)
\bbtitle{Software and Cyberinfrastructure for Astronomy IV}.
\bsertitle{Society of Photo-Optical Instrumentation Engineers (SPIE) Conference Series},
vol. \bseriesno{9913},
p. \bfpage{99133}
(\byear{2016})
\end{bchapter}
\endbibitem
%%% 51
\bibitem[\protect\citeauthoryear{{Twicken} et~al.}{2010}]{SAP_Twickens_2010}
\begin{bchapter}
\bauthor{\bsnm{{Twicken}}, \binits{J.D.}} \betal:
\bctitle{{Photometric analysis in the Kepler Science Operations Center pipeline}}.
In: \beditor{\bsnm{{Radziwill}}, \binits{N.M.}},
\beditor{\bsnm{{Bridger}}, \binits{A.}} (eds.)
\bbtitle{Software and Cyberinfrastructure for Astronomy}.
\bsertitle{Proc.\ SPIE},
vol. \bseriesno{7740},
p. \bfpage{774023}
(\byear{2010})
\end{bchapter}
\endbibitem

%%% 52
\bibitem[\protect\citeauthoryear{{Vanderburg} et~al.}{2019}]{Vanderburg2019}
\begin{barticle}
\bauthor{\bsnm{{Vanderburg}}, \binits{A.}} \betal:
\batitle{{TESS Spots a Compact System of Super-Earths around the Naked-eye Star HR 858}}.
\bjtitle{Astrophys. J. Lett.}
\bvolume{881}(\bissue{1}),
\bfpage{19}
(\byear{2019})
\end{barticle}
\endbibitem

%%% 53
\bibitem[\protect\citeauthoryear{{Barber} et~al.}{2024}]{Barber}
\begin{barticle}
\bauthor{\bsnm{{Barber}}, \binits{M.G.}} \betal:
\batitle{{TESS Investigation—Demographics of Young Exoplanets (TI-DYE). II. A Second Giant Planet in the 17 Myr System HIP 67522}}.
\bjtitle{Astrophys. J. Lett.}
\bvolume{973}(\bissue{1}),
\bfpage{L30}
(\byear{2024})
\end{barticle}
\endbibitem


%%% 54
\bibitem[\protect\citeauthoryear{{Collins}}{2019}]{collins:2019}
\begin{bchapter}
\bauthor{\bsnm{{Collins}}, \binits{K.}}:
\bctitle{{TESS Follow-up Observing Program Working Group (TFOP WG) Sub Group 1 (SG1): Ground-based Time-series Photometry}}.
In: \bbtitle{American Astronomical Society Meeting Abstracts \#233}.
\bsertitle{AAS Meeting Abstr.},
vol. \bseriesno{233},
pp. \bfpage{140}--\blpage{05}
(\byear{2019})
\end{bchapter}
\endbibitem

%%% 55
\bibitem[\protect\citeauthoryear{{Brown} et~al.}{2013}]{Brown:2013}
\begin{barticle}
\bauthor{\bsnm{{Brown}}, \binits{T.M.}} \betal:
\batitle{{Las Cumbres Observatory Global Telescope Network}}.
\bjtitle{Publ. Astron. Soc. Pac.}
\bvolume{125},
\bfpage{1031}
(\byear{2013})
\end{barticle}
\endbibitem

%%% 56
\bibitem[\protect\citeauthoryear{{Jensen}}{2013}]{Jensen:2013}
\begin{botherref}
\oauthor{\bsnm{{Jensen}}, \binits{E.}}:
{Tapir: A web interface for transit/eclipse observability}.
Astrophysics Source Code Library
(\byear{2013})
\end{botherref}
\endbibitem

%%% 57
\bibitem[\protect\citeauthoryear{{Narita} et~al.}{2020}]{Narita:2020}
\begin{bchapter}
\bauthor{\bsnm{{Narita}}, \binits{N.}} \betal:
\bctitle{{MuSCAT3: a 4-color simultaneous camera for the 2m Faulkes Telescope North}}.
In: \bbtitle{Society of Photo-Optical Instrumentation Engineers (SPIE) Conference Series}.
\bsertitle{Proc.\ SPIE},
vol. \bseriesno{11447},
p. \bfpage{114475}
(\byear{2020})
\end{bchapter}
\endbibitem

%%% 58
\bibitem[\protect\citeauthoryear{{McCully} et~al.}{2018}]{McCully:2018}
\begin{bchapter}
\bauthor{\bsnm{{McCully}}, \binits{C.}} \betal:
\bctitle{{Real-time processing of the imaging data from the network of Las Cumbres Observatory Telescopes using BANZAI}}.
In: \bbtitle{Proceedings of SPIE}.
\bsertitle{Proc.\ SPIE},
vol. \bseriesno{10707},
p. \bfpage{107070}
(\byear{2018})
\end{bchapter}
\endbibitem

%%% 59
\bibitem[\protect\citeauthoryear{{Collins} et~al.}{2017}]{Collins:2017}
\begin{barticle}
\bauthor{\bsnm{{Collins}}, \binits{K.A.}},
\bauthor{\bsnm{{Kielkopf}}, \binits{J.F.}},
\bauthor{\bsnm{{Stassun}}, \binits{K.G.}},
\bauthor{\bsnm{{Hessman}}, \binits{F.V.}}:
\batitle{{AstroImageJ: Image Processing and Photometric Extraction for Ultra-precise Astronomical Light Curves}}.
\bjtitle{Astron. J.}
\bvolume{153},
\bfpage{77}
(\byear{2017})
\end{barticle}
\endbibitem

%%% 60
\bibitem[\protect\citeauthoryear{{Benz} et~al.}{2021}]{CHEOPS}
\begin{barticle}
\bauthor{\bsnm{{Benz}}, \binits{W.}} \betal:
\batitle{{The CHEOPS mission}}.
\bjtitle{Exp. Astron.}
\bvolume{51}(\bissue{1}),
\bfpage{109}--\blpage{151}
(\byear{2021})
\end{barticle}
\endbibitem

%%% 61
\bibitem[\protect\citeauthoryear{Osborn et~al.}{2022}]{Osborn2022}
\begin{barticle}
\bauthor{\bsnm{{Osborn}}, \binits{H.P.}} \betal:
\batitle{{Uncovering the true periods of the young sub-Neptunes orbiting TOI-2076}}.
\bjtitle{Astron. Astrophys.}
\bvolume{664},
\bfpage{1}--\blpage{17}
(\byear{2022})
\end{barticle}
\endbibitem

%%% 62
\bibitem[\protect\citeauthoryear{{Hoyer} et~al.}{2020}]{CHEOPS_DRP_2020}
\begin{barticle}
\bauthor{\bsnm{{Hoyer}}, \binits{S.}} \betal:
\batitle{{Expected performances of the Characterising Exoplanet Satellite (CHEOPS). III. Data reduction pipeline: architecture and simulated performances}}.
\bjtitle{Astron. Astrophys.}
\bvolume{635},
\bfpage{24}
(\byear{2020})
\end{barticle}
\endbibitem

%%% 63
\bibitem[\protect\citeauthoryear{{Mandel} and {Agol}}{2002}]{Mandel2002}
\begin{barticle}
\bauthor{\bsnm{{Mandel}}, \binits{K.}},
\bauthor{\bsnm{{Agol}}, \binits{E.}}:
\batitle{{Analytic Light Curves for Planetary Transit Searches}}.
\bjtitle{Astrophys. J.}
\bvolume{580}(\bissue{2}),
\bfpage{171}--\blpage{175}
(\byear{2002})
\end{barticle}
\endbibitem

%%% 64
\bibitem[\protect\citeauthoryear{{Kreidberg}}{2015}]{Kreidberg2015}
\begin{barticle}
\bauthor{\bsnm{{Kreidberg}}, \binits{L.}}:
\batitle{{batman: BAsic Transit Model cAlculatioN in Python}}.
\bjtitle{Publ. Astron. Soc. Pac.}
\bvolume{127}(\bissue{957}),
\bfpage{1161}
(\byear{2015})
\end{barticle}
\endbibitem

%%% 65
\bibitem[\protect\citeauthoryear{{Espinoza} et~al.}{2019}]{Espinoza2019}
\begin{barticle}
\bauthor{\bsnm{{Espinoza}}, \binits{N.}},
\bauthor{\bsnm{{Kossakowski}}, \binits{D.}},
\bauthor{\bsnm{{Brahm}}, \binits{R.}}:
\batitle{{juliet: a versatile modelling tool for transiting and non-transiting exoplanetary systems}}.
\bjtitle{Mon. Not. R. Astron. Soc.}
\bvolume{490}(\bissue{2}),
\bfpage{2262}--\blpage{2283}
(\byear{2019})
\end{barticle}
\endbibitem

%%% 66
\bibitem[\protect\citeauthoryear{{Damasso} et~al.}{2024}]{Damasso2024}
\begin{barticle}
\bauthor{\bsnm{{Damasso}}, \binits{M.}} \betal:
\batitle{{The GAPS Programme at TNG: LIX. Characterisation study of the {\ensuremath{\sim}}300 Myr-old multi-planetary system orbiting the star BD+40 2790 (TOI-2076)}}.
\bjtitle{Astron. Astrophys.}
\bvolume{690},
\bfpage{235}
(\byear{2024})
\end{barticle}
\endbibitem

%%% 67
\bibitem[\protect\citeauthoryear{{Polanski} et~al.}{2024}]{Polanski2024}
\begin{barticle}
\bauthor{\bsnm{{Polanski}}, \binits{A.S.}} \betal:
\batitle{{The TESS-Keck Survey. XX. 15 New TESS Planets and a Uniform RV Analysis of All Survey Targets}}.
\bjtitle{Astrophys. J. Suppl. Ser.}
\bvolume{272}(\bissue{2}),
\bfpage{32}
(\byear{2024})
\end{barticle}
\endbibitem

%%% 68
\bibitem[\protect\citeauthoryear{{Grunblatt} et~al.}{2015}]{Grunblatt2015}
\begin{barticle}
\bauthor{\bsnm{{Grunblatt}}, \binits{S.K.}},
\bauthor{\bsnm{{Howard}}, \binits{A.W.}},
\bauthor{\bsnm{{Haywood}}, \binits{R.D.}}:
\batitle{{Determining the Mass of Kepler-78b with Nonparametric Gaussian Process Estimation}}.
\bjtitle{Astrophys. J.}
\bvolume{808},
\bfpage{127}
(\byear{2015})
\end{barticle}
\endbibitem

%%% 69
\bibitem[\protect\citeauthoryear{{Haywood} et~al.}{2014}]{Haywood}
\begin{barticle}
\bauthor{\bsnm{{Haywood}}, \binits{R.D.}} \betal:
\batitle{{Planets and stellar activity: hide and seek in the CoRoT-7 system}}.
\bjtitle{Mon. Not. R. Astron. Soc.}
\bvolume{443},
\bfpage{2517}--\blpage{2531}
(\byear{2014})
\end{barticle}
\endbibitem

%%% 70
\bibitem[\protect\citeauthoryear{{Aigrain} et~al.}{2012}]{Aigrain2012}
\begin{barticle}
\bauthor{\bsnm{{Aigrain}}, \binits{S.}},
\bauthor{\bsnm{{Pont}}, \binits{F.}},
\bauthor{\bsnm{{Zucker}}, \binits{S.}}:
\batitle{{A simple method to estimate radial velocity variations due to stellar activity using photometry}}.
\bjtitle{Mon. Not. R. Astron. Soc.}
\bvolume{419}(\bissue{4}),
\bfpage{3147}--\blpage{3158}
(\byear{2012})
\end{barticle}
\endbibitem

%%% 71
\bibitem[\protect\citeauthoryear{{Dai} et~al.}{2019}]{Dai2019}
\begin{barticle}
\bauthor{\bsnm{{Dai}}, \binits{F.}},
\bauthor{\bsnm{{Masuda}}, \binits{K.}},
\bauthor{\bsnm{{Winn}}, \binits{J.N.}},
\bauthor{\bsnm{{Zeng}}, \binits{L.}}:
\batitle{{Homogeneous Analysis of Hot Earths: Masses, Sizes, and Compositions}}.
\bjtitle{Astrophys. J.}
\bvolume{883}(\bissue{1}),
\bfpage{79}
(\byear{2019})
\end{barticle}
\endbibitem

%%% 72
\bibitem[\protect\citeauthoryear{{Zechmeister} et~al.}{2018}]{SERVAL}
\begin{barticle}
\bauthor{\bsnm{{Zechmeister}}, \binits{M.}} \betal:
\batitle{{Spectrum radial velocity analyser (SERVAL). High-precision radial velocities and two alternative spectral indicators}}.
\bjtitle{Astron. Astrophys.}
\bvolume{609},
\bfpage{12}
(\byear{2018})
\end{barticle}
\endbibitem

%%% 73
\bibitem[\protect\citeauthoryear{{Artigau} et~al.}{2022}]{HARPS_lbl}
\begin{barticle}
\bauthor{\bsnm{{Artigau}}, \binits{{\'E}.}} \betal:
\batitle{{Line-by-line Velocity Measurements: an Outlier-resistant Method for Precision Velocimetry}}.
\bjtitle{Astron. J.}
\bvolume{164}(\bissue{3}),
\bfpage{84}
(\byear{2022})
\end{barticle}
\endbibitem

%%% 74
\bibitem[\protect\citeauthoryear{{Speagle}}{2020}]{dynesty2020}
\begin{barticle}
\bauthor{\bsnm{{Speagle}}, \binits{J.S.}}:
\batitle{{DYNESTY: a dynamic nested sampling package for estimating Bayesian posteriors and evidences}}.
\bjtitle{Mon. Not. R. Astron. Soc.}
\bvolume{493}(\bissue{3}),
\bfpage{3132}--\blpage{3158}
(\byear{2020})
\end{barticle}
\endbibitem

%%% 75
\bibitem[\protect\citeauthoryear{{Blunt} et~al.}{2023}]{Blunt}
\begin{barticle}
\bauthor{\bsnm{{Blunt}}, \binits{S.}} \betal:
\batitle{{Overfitting Affects the Reliability of Radial Velocity Mass Estimates of the V1298 Tau Planets}}.
\bjtitle{Astron. J.}
\bvolume{166}(\bissue{2}),
\bfpage{62}
(\byear{2023})
\end{barticle}
\endbibitem

%%% 76
\bibitem[\protect\citeauthoryear{{Masuda} et~al.}{2024}]{Masuda2024}
\begin{barticle}
\bauthor{\bsnm{{Masuda}}, \binits{K.}} \betal:
\batitle{{A Fourth Planet in the Kepler-51 System Revealed by Transit Timing Variations}}.
\bjtitle{Astron. J.}
\bvolume{168}(\bissue{6}),
\bfpage{294}
(\byear{2024})
\end{barticle}
\endbibitem


%%% 77
\bibitem[\protect\citeauthoryear{Phan et~al.}{2019}]{numpyro2019}
\begin{botherref}
\oauthor{\bsnm{Phan}, \binits{D.}} \betal:
{Composable effects for flexible and accelerated probabilistic programming in numpyro}.
arXiv e-prints,
1912.11554
(\byear{2019})
\end{botherref}
\endbibitem

%%% 78
\bibitem[\protect\citeauthoryear{Brooks and Gelman}{1998}]{Brooks1998}
\begin{barticle}
\bauthor{\bsnm{Brooks}, \binits{S.P.}},
\bauthor{\bsnm{Gelman}, \binits{A.}}:
\batitle{{General methods for monitoring convergence of iterative simulations}}.
\bjtitle{J. Comput. Graph. Stat.}
\bvolume{7}(\bissue{4}),
\bfpage{434}--\blpage{455}
(\byear{1998})
\end{barticle}
\endbibitem

%%% 79
\bibitem[\protect\citeauthoryear{{Lithwick} et~al.}{2012}]{Lithwick_ttv}
\begin{barticle}
\bauthor{\bsnm{{Lithwick}}, \binits{Y.}},
\bauthor{\bsnm{{Xie}}, \binits{J.}},
\bauthor{\bsnm{{Wu}}, \binits{Y.}}:
\batitle{{Extracting Planet Mass and Eccentricity from TTV Data}}.
\bjtitle{Astrophys. J.}
\bvolume{761}(\bissue{2}),
\bfpage{122}
(\byear{2012})
\end{barticle}
\endbibitem

%%% 80
\bibitem[\protect\citeauthoryear{Hadden and Lithwick}{2017}]{Hadden2017}
\begin{barticle}
\bauthor{\bsnm{Hadden}, \binits{S.}},
\bauthor{\bsnm{Lithwick}, \binits{Y.}}:
\batitle{{Kepler Planet Masses and Eccentricities from TTV Analysis}}.
\bjtitle{Astron. J.}
\bvolume{154}(\bissue{1}),
\bfpage{5}
(\byear{2017})
\end{barticle}
\endbibitem

%%% 81
\bibitem[\protect\citeauthoryear{Leleu et~al.}{2023}]{Leleu2023}
\begin{barticle}
\bauthor{\bsnm{Leleu}, \binits{A.}} \betal:
\batitle{{Removing biases on the density of sub-Neptunes characterised via transit timing variations: Update on the mass-radius relationship of 34 Kepler planets}}.
\bjtitle{Astron. Astrophys.}
\bvolume{669},
\bfpage{1}--\blpage{19}
(\byear{2023})
\end{barticle}
\endbibitem

%%% 82
\bibitem[\protect\citeauthoryear{{Wang} and {Liu}}{2024}]{Wang2024}
\begin{barticle}
\bauthor{\bsnm{{Wang}}, \binits{M.-T.}},
\bauthor{\bsnm{{Liu}}, \binits{H.-G.}}:
\batitle{{Photo-dynamical Analysis of Circumbinary Multi-planet System TOI-1338: A Fully Coplanar Configuration with a Puffy Planet}}.
\bjtitle{Astron. J.}
\bvolume{168}(\bissue{1}),
\bfpage{31}
(\byear{2024})
\end{barticle}
\endbibitem

%%% 83
\bibitem[\protect\citeauthoryear{{Jontof-Hutter} et~al.}{2016}]{JontofHutter2016}
\begin{barticle}
\bauthor{\bsnm{{Jontof-Hutter}}, \binits{D.}} \betal:
\batitle{{Secure Mass Measurements from Transit Timing: 10 Kepler Exoplanets between 3 and 8 M$_{{\ensuremath{\oplus}}}$ with Diverse Densities and Incident Fluxes}}.
\bjtitle{Astrophys. J.}
\bvolume{820}(\bissue{1}),
\bfpage{39}
(\byear{2016})
\end{barticle}
\endbibitem

%%% 84
\bibitem[\protect\citeauthoryear{{Rein} and {Tamayo}}{2015}]{whfast2015}
\begin{barticle}
\bauthor{\bsnm{{Rein}}, \binits{H.}},
\bauthor{\bsnm{{Tamayo}}, \binits{D.}}:
\batitle{{WHFAST: a fast and unbiased implementation of a symplectic Wisdom-Holman integrator for long-term gravitational simulations}}.
\bjtitle{Mon. Not. R. Astron. Soc.}
\bvolume{452}(\bissue{1}),
\bfpage{376}--\blpage{388}
(\byear{2015})
\end{barticle}
\endbibitem

%%% 85
\bibitem[\protect\citeauthoryear{{Rein} and {Liu}}{2012}]{ReinLiu2012}
\begin{barticle}
\bauthor{\bsnm{{Rein}}, \binits{H.}},
\bauthor{\bsnm{{Liu}}, \binits{S.-F.}}:
\batitle{{REBOUND: an open-source multi-purpose N-body code for collisional dynamics}}.
\bjtitle{Astron. Astrophys.}
\bvolume{537},
\bfpage{128}
(\byear{2012})
\end{barticle}
\endbibitem

%%% 86
\bibitem[\protect\citeauthoryear{{Masuda} et~al.}{2013}]{Masuda2013}
\begin{barticle}
\bauthor{\bsnm{{Masuda}}, \binits{K.}} \betal:
\batitle{{Characterization of the KOI-94 System with Transit Timing Variation Analysis: Implication for the Planet-Planet Eclipse}}.
\bjtitle{Astrophys. J.}
\bvolume{778},
\bfpage{185}
(\byear{2013})
\end{barticle}
\endbibitem

%%% 87
\bibitem[\protect\citeauthoryear{{Foreman-Mackey} et~al.}{2013}]{emcee}
\begin{barticle}
\bauthor{\bsnm{{Foreman-Mackey}}, \binits{D.}} \betal:
\batitle{{emcee: The MCMC Hammer}}.
\bjtitle{Publ. Astron. Soc. Pac.}
\bvolume{125},
\bfpage{306}--\blpage{312}
(\byear{2013})
\end{barticle}
\endbibitem

%%% 88
\bibitem[\protect\citeauthoryear{{Lee}}{2019}]{Lee2019}
\begin{barticle}
\bauthor{\bsnm{{Lee}}, \binits{E.J.}}:
\batitle{{The Boundary between Gas-rich and Gas-poor Planets}}.
\bjtitle{Astrophys. J.}
\bvolume{878}(\bissue{1}),
\bfpage{36}
(\byear{2019})
\end{barticle}
\endbibitem

%%% 89
\bibitem[\protect\citeauthoryear{{Gu} and {Chen}}{2023}]{Gu+Chen23}
\begin{barticle}
\bauthor{\bsnm{{Gu}}, \binits{P.-G.}},
\bauthor{\bsnm{{Chen}}, \binits{H.}}:
\batitle{{Deuterium Escape on Photoevaporating Sub-Neptunes}}.
\bjtitle{Astrophys. J. Lett.}
\bvolume{953}(\bissue{2}),
\bfpage{27}
(\byear{2023})
\end{barticle}
\endbibitem

%%% 90
\bibitem[\protect\citeauthoryear{{Malsky} and {Rogers}}{2020}]{Malsky20}
\begin{barticle}
\bauthor{\bsnm{{Malsky}}, \binits{I.}},
\bauthor{\bsnm{{Rogers}}, \binits{L.A.}}:
\batitle{{Coupled Thermal and Compositional Evolution of Photoevaporating Planet Envelopes}}.
\bjtitle{Astrophys. J.}
\bvolume{896}(\bissue{1}),
\bfpage{48}
(\byear{2020})
\end{barticle}
\endbibitem

%%% 91
\bibitem[\protect\citeauthoryear{{Watson} et~al.}{1981}]{Watson1981}
\begin{barticle}
\bauthor{\bsnm{{Watson}}, \binits{A.J.}},
\bauthor{\bsnm{{Donahue}}, \binits{T.M.}},
\bauthor{\bsnm{{Walker}}, \binits{J.C.G.}}:
\batitle{{The dynamics of a rapidly escaping atmosphere: Applications to the evolution of Earth and Venus}}.
\bjtitle{Icarus}
\bvolume{48}(\bissue{2}),
\bfpage{150}--\blpage{166}
(\byear{1981})
\end{barticle}
\endbibitem

%%% 92
\bibitem[\protect\citeauthoryear{{Erkaev} et~al.}{2014}]{ErkeavEt2014}
\begin{barticle}
\bauthor{\bsnm{{Erkaev}}, \binits{N.V.}} \betal:
\batitle{{Escape of the martian protoatmosphere and initial water inventory}}.
\bjtitle{Planet. Space Sci.}
\bvolume{98},
\bfpage{106}--\blpage{119}
(\byear{2014})
\end{barticle}
\endbibitem

%%% 93
\bibitem[\protect\citeauthoryear{{Murray-Clay} et~al.}{2009}]{Murray-Clay2009}
\begin{barticle}
\bauthor{\bsnm{{Murray-Clay}}, \binits{R.A.}},
\bauthor{\bsnm{{Chiang}}, \binits{E.I.}},
\bauthor{\bsnm{{Murray}}, \binits{N.}}:
\batitle{{Atmospheric Escape From Hot Jupiters}}.
\bjtitle{Astrophys. J.}
\bvolume{693}(\bissue{1}),
\bfpage{23}--\blpage{42}
(\byear{2009})
\end{barticle}
\endbibitem

%%% 94
\bibitem[\protect\citeauthoryear{Jackson et~al.}{2012}]{Jackson2012}
\begin{barticle}
\bauthor{\bsnm{Jackson}, \binits{A.P.}},
\bauthor{\bsnm{Davis}, \binits{T.A.}},
\bauthor{\bsnm{Wheatley}, \binits{P.J.}}:
\batitle{{The coronal X-ray-age relation and its implications for the evaporation of exoplanets}}.
\bjtitle{Mon. Not. R. Astron. Soc.}
\bvolume{422}(\bissue{3}),
\bfpage{2024}--\blpage{2043}
(\byear{2012})
\end{barticle}
\endbibitem

%%% 95
\bibitem[\protect\citeauthoryear{{Valencia} et~al.}{2010}]{Valencia2010}
\begin{barticle}
\bauthor{\bsnm{{Valencia}}, \binits{D.}},
\bauthor{\bsnm{{Ikoma}}, \binits{M.}},
\bauthor{\bsnm{{Guillot}}, \binits{T.}},
\bauthor{\bsnm{{Nettelmann}}, \binits{N.}}:
\batitle{{Composition and fate of short-period super-Earths. The case of CoRoT-7b}}.
\bjtitle{Astron. Astrophys.}
\bvolume{516},
\bfpage{20}
(\byear{2010})
\end{barticle}
\endbibitem

%%% 96
\bibitem[\protect\citeauthoryear{Tian and Ida}{2015}]{tian2015}
\begin{barticle}
\bauthor{\bsnm{Tian}, \binits{F.}},
\bauthor{\bsnm{Ida}, \binits{S.}}:
\batitle{{Water contents of earth-mass planets around M dwarfs}}.
\bjtitle{Nat. Geosci.}
\bvolume{8}(\bissue{3}),
\bfpage{177}--\blpage{180}
(\byear{2015})
\end{barticle}
\endbibitem

%%% 97
\bibitem[\protect\citeauthoryear{Owen and Schlichting}{2024}]{Owen2024}
\begin{barticle}
\bauthor{\bsnm{Owen}, \binits{J.E.}},
\bauthor{\bsnm{Schlichting}, \binits{H.E.}}:
\batitle{{Mapping out the parameter space for photoevaporation and core-powered mass-loss}}.
\bjtitle{Mon. Not. R. Astron. Soc.}
\bvolume{528}(\bissue{2}),
\bfpage{1615}--\blpage{1629}
(\byear{2024})
\end{barticle}
\endbibitem

%%% 98
\bibitem[\protect\citeauthoryear{{Rogers} et~al.}{2024}]{rogers2024}
\begin{barticle}
\bauthor{\bsnm{{Rogers}}, \binits{J.G.}} \betal:
\batitle{{Under the light of a new star: evolution of planetary atmospheres through protoplanetary disc dispersal and boil-off}}.
\bjtitle{Mon. Not. R. Astron. Soc.}
\bvolume{529}(\bissue{3}),
\bfpage{2716}--\blpage{2733}
(\byear{2024})
\end{barticle}
\endbibitem

%%% 99
\bibitem[\protect\citeauthoryear{{Tang} et~al.}{2025}]{Tang2025}
\begin{barticle}
\bauthor{\bsnm{{Tang}}, \binits{Y.}},
\bauthor{\bsnm{{Fortney}}, \binits{J.J.}},
\bauthor{\bsnm{{Murray-Clay}}, \binits{R.}},
\bauthor{\bsnm{{Broome}}, \binits{M.}}:
\batitle{{Understanding the Origins of Super-puff Planets: A New Mass-loss Regime Coupled to Planetary Evolution}}.
\bjtitle{Astrophys. J.}
\bvolume{995}(\bissue{1}),
\bfpage{20}
(\byear{2025})
\end{barticle}
\endbibitem

%%% 100
\bibitem[\protect\citeauthoryear{King and Wheatley}{2021}]{King2021}
\begin{barticle}
\bauthor{\bsnm{King}, \binits{G.W.}},
\bauthor{\bsnm{Wheatley}, \binits{P.J.}}:
\batitle{{EUV irradiation of exoplanet atmospheres occurs on Gyr time-scales}}.
\bjtitle{Mon. Not. R. Astron. Soc. Lett.}
\bvolume{501}(\bissue{1}),
\bfpage{28}--\blpage{32}
(\byear{2021})
\end{barticle}
\endbibitem

%%% 101
\bibitem[\protect\citeauthoryear{{Tian}}{2009}]{Tian2009}
\begin{barticle}
\bauthor{\bsnm{{Tian}}, \binits{F.}}:
\batitle{{Thermal Escape from Super Earth Atmospheres in the Habitable Zones of M Stars}}.
\bjtitle{Astrophys. J.}
\bvolume{703}(\bissue{1}),
\bfpage{905}--\blpage{909}
(\byear{2009})
\end{barticle}
\endbibitem


%%% 103
\bibitem[\protect\citeauthoryear{{Masuda}}{2025}]{2025ascl.soft05006M}
\begin{botherref}
\oauthor{\bsnm{{Masuda}}, \binits{K.}}:
{jnkepler: Differentiable N-body Model for Multi-planet Systems}.
\textit{Astrophysics Source Code Library}
ascl:2505.006
(\byear{2025})
\end{botherref}
\endbibitem

%%% 104
\bibitem[\protect\citeauthoryear{{Tayar}, {Claytor}, {Huber} and {van Saders}}{2022}]{Tayar2022}
\begin{barticle}
\bauthor{\bsnm{{Tayar}}, \binits{J.}},
\bauthor{\bsnm{{Claytor}}, \binits{Z.R.}},
\bauthor{\bsnm{{Huber}}, \binits{D.}},
\bauthor{\bsnm{{van Saders}}, \binits{J.}}:
\batitle{{A Guide to Realistic Uncertainties on the Fundamental Properties of Solar-type Exoplanet Host Stars}}.
\bjtitle{Astrophys. J.}
\bvolume{927}(\bissue{1}),
\bfpage{31}
(\byear{2022})
\end{barticle}
\endbibitem


%%% 105 (you included after 104)
\bibitem[\protect\citeauthoryear{{Wang}}{2025}]{Wang2025figshare}
\begin{barticle}
\bauthor{\bsnm{{Wang}}, \binits{M.}}
\batitle{{An Adolescent, Near-Resonant Planetary System Near the End of Photoevaporation}}.
\bjtitle{figshare. Dataset}
(\byear{2025})
\doiurl{10.6084/m9.figshare.29815580}
\end{barticle}
\endbibitem

\bibitem[\protect\citeauthoryear{{Damasso} et~al.}{2024}]{Damasso2024Vizier}
\begin{barticle}
\bauthor{\bsnm{{Damasso}}, \binits{M.}} \betal:
\batitle{{Spectroscopic data of BD+40 2790 (TOI-2076)}}.
\bjtitle{VizieR Online Data Catalog}.
\bpublisher{Centre de Donn{\'e}es astronomiques de Strasbourg (CDS)}
(\byear{2024})
\doiurl{https://doi.org/10.26093/cds/vizier.36900235}
\end{barticle}
\endbibitem

\bibitem[\protect\citeauthoryear{{MAST Team}}{2021}]{MASTtesslc}
\begin{barticle}
\bauthor{\bsnm{{MAST Team}}}:
\batitle{{TESS light curves -- all sectors}}.
\bjtitle{Mikulski Archive for Space Telescopes (MAST)}
(\byear{2021})
\doiurl{https://doi.org/10.17909/T9-NMC8-F686}
\end{barticle}

\end{thebibliography}
\end{document}